\documentclass[aps,prl,reprint,groupedaddress,showpacs,amsmath,amssymb,twocolumn,floatfix]{revtex4-1}
\usepackage{graphicx}
\usepackage{bm}
\usepackage{color}
\usepackage[normalem]{ulem}
\usepackage{amsmath}
\usepackage{amssymb}
\renewcommand{\vec}[1]{\boldsymbol{#1}}
\newcommand{\pdagger}{{\phantom{\dagger}}}

\newcommand{\red}{\textcolor{red}}

\newcommand{\wse}{WSe$_2$}

\newcommand{\WSe}{WSe$_2$}
\newcommand{\Rxx}{$R_{xx}$}
\newcommand{\RHP}{$R_{xx}^{HP}$}

\newcommand{\DRxx}{$\Delta R_{xx}$}

\newcommand{\equref}[1]{Eq.~(\ref{#1})}
\newcommand{\equsref}[2]{Eqs.~(\ref{#1}) and (\ref{#2})}

\newcommand{\vecB}{$\vec{B}$}
\newcommand{\Bperp}{$B_{\perp}$}
\newcommand{\Bpara}{$B_{\parallel}$}

\usepackage[colorlinks=true]{hyperref}  
\hypersetup{
    bookmarks=true,         
    unicode=false,          
    pdftoolbar=true,        
    pdfmenubar=true,        
    pdffitwindow=false,     
    pdfstartview={FitH},    
    pdftitle={Low-energy collective modes in MATBG},    
    pdfauthor={Morissette et al.},     
    pdfsubject={},   
    pdfcreator={},   
    pdfproducer={}, 
    pdfkeywords={} {} {}, 
    pdfnewwindow=true,      
    colorlinks=true,       
    linkcolor=magenta, 
    citecolor=blue,        
    filecolor=magenta,      
    urlcolor=blue           
} 
\usepackage{cleveref}

\begin{document}

\title{Electron spin resonance and collective excitations  in magic-angle twisted bilayer graphene}

\author{Erin Morissette$^{1}$}
\author{Jiang-Xiazi Lin$^{1}$}
\author{Dihao Sun$^{1}$}
\author{Liangji Zhang$^{2}$}
\author{Song Liu$^{3}$}
\author{Daniel Rhodes$^{3}$}
\author{Kenji Watanabe$^{4}$}
\author{Takashi Taniguchi$^{5}$}
\author{James Hone$^{3}$}
\author{Johannes~Pollanen$^{2}$}
\author{Mathias S. Scheurer$^{6}$}
\author{Michael Lilly$^{7}$}
\author{Andrew Mounce$^{7*}$}
\author{J.I.A. Li$^{1}$}
\email{amounce@sandia.gov}
\email{jia\_li@brown.edu}

\affiliation{$^{1}$Department of Physics, Brown University, Providence, RI 02912, USA}
\affiliation{$^{2}$Department of Physics and Astronomy, Michigan State University, East Lansing, MI 48824 USA}
\affiliation{$^{3}$Department of Mechanical Engineering, Columbia University, New York, NY 10027, USA}
\affiliation{$^{4}$Research Center for Functional Materials, National Institute for Materials Science, 1-1 Namiki, Tsukuba 305-0044, Japan}
\affiliation{$^{5}$International Center for Materials Nanoarchitectonics,
National Institute for Materials Science,  1-1 Namiki, Tsukuba 305-0044, Japan}
\affiliation{$^{6}$ Institute for Theoretical Physics, University of Innsbruck, Innsbruck, A-6020, Austria}
\affiliation{$^{7}$ Center for Integrated Nanotechnologies, Sandia National Laboratories, Albuquerque, New Mexico 87123, USA}

\date{\today}

\maketitle

\textbf{In a strongly correlated system, collective excitations contain key information regarding the electronic order of the underlying ground state. An abundance of collective modes in the spin and valley isospin channels of magic-angle graphene moir\'e bands has been alluded to by a series of recent experiments ~\cite{Saito2021pomeranchuk,Rozen2021pomeranchuk,Liu2022DtTLG}. 
However, direct observation of collective excitations has remained elusive due to the lack of a spin probe.
In this work, we use a resistively-detected electron spin resonance technique to look for low-energy collective excitations in magic-angle twisted bilayer graphene. We report direct observation of collective modes in the form of microwave-induced resonance near half filling of the moir\'e flatbands. The frequency-magnetic field dependence of these resonance modes sheds light onto the nature of intervalley spin coupling, allowing us to extract parameters such as intervalley exchange interaction and spin stiffness.
Two independent observations testify that the generation and detection of the microwave resonance relies on the strong correlation within the flat moir\'e energy band.  First, the onset of robust resonance response coincides with the spontaneous flavor polarization at half moir\'e filling, and remains absent in the density range where the underlying Fermi surface is isospin unpolarized. Second, we performed the same resonance measurement on graphene monolayer and bilayer samples, including twisted bilayer with a large twist angle, where flatband physics is absent. We observe no indication of resonance response in these samples across a large range of carrier density, microwave frequency and power. A natural explanation is that the resonance response near the magic angle originates from ``Dirac revivals'' and the resulting isospin order ~\cite{Zondiner2020cascade,Park2021flavour,Saito2021pomeranchuk,Rozen2021pomeranchuk}. }

\begin{figure*}
\includegraphics[width=0.9\linewidth]{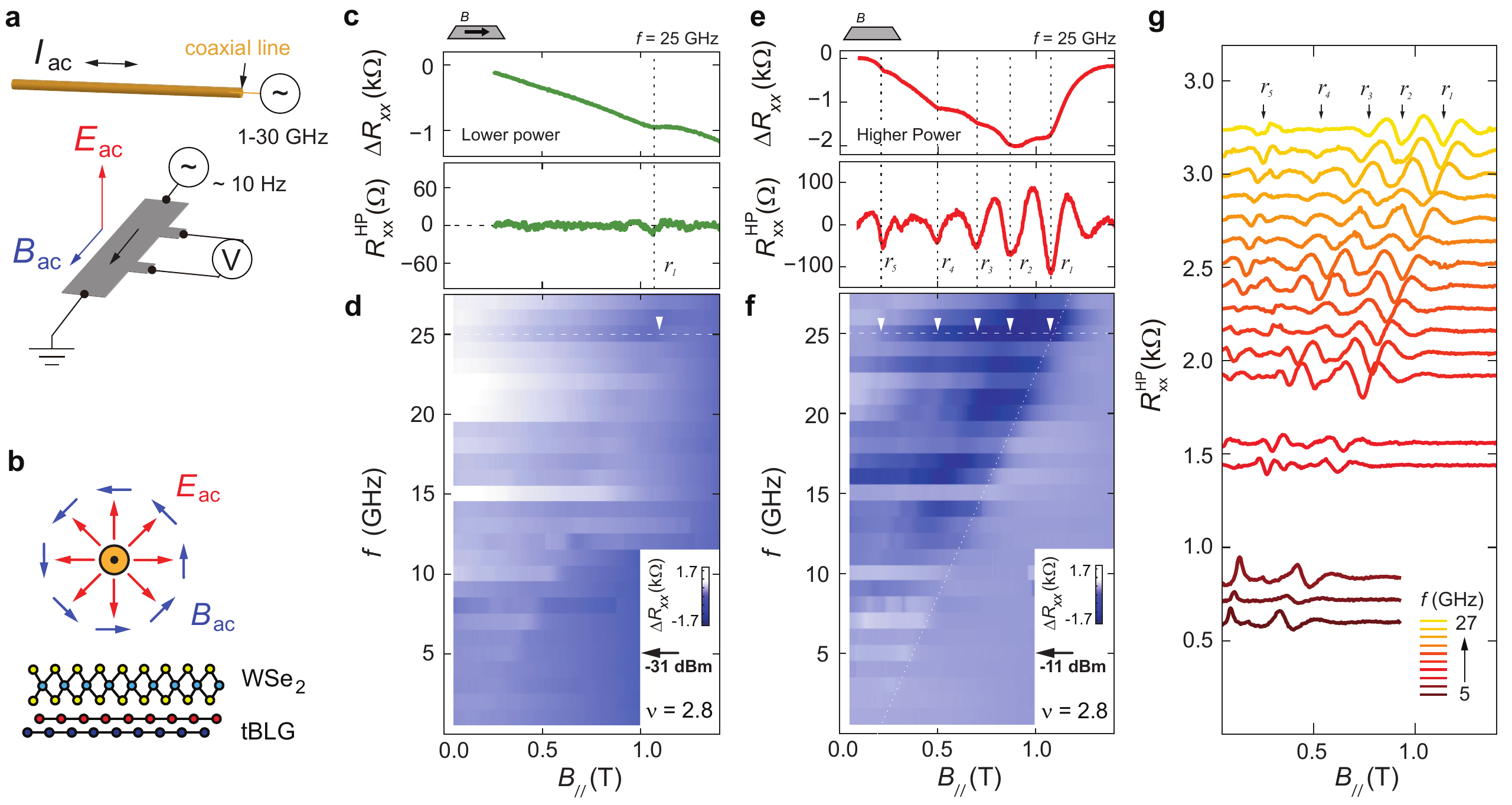} 
\caption{\label{fig1}{\bf{Microwave-induced transport response in magic-angle twisted bilayer graphene.}}
(a) A schematic of the device measurement setup, including the microwave coaxial line, $I_{ac}$, proximally above the Hall bar channel. Microwave signal with frequency of $1-30$ GHz is passed through the coaxial line. Longitudinal resistance \Rxx\ is measured from the sample with a small current bias of $\sim 5$ nA at a near-DC frequency of $17$ Hz. 
(b) According to the geometry of the setup, the magnetic (electric) component of the microwave field, $B_{ac}$ ($E_{ac}$), is aligned parallel (perpendicular) to the graphene sheets. 
(c-f) The influence of microwave radiation on the transport response of the sample measured at $\nu=2.8$. \DRxx\ is the change in longitudinal resistance compared to $B=0$.  \RHP\ is obtained by applying a high-pass filter (HPF) to eliminate the slow-varying background in \DRxx.  \DRxx\ and \RHP\ measured with (c-d) low microwave power of $0.8$ dBm and (e-f) high microwave power of $16.8$ dBm. 
(c) and (e) \DRxx\ (top panel)  and \RHP\ (bottom panel) as a function of an in-plane magnetic field \Bpara. (d) and (f) \DRxx\ as a function of an in-plane magnetic field \Bpara\ and microwave frequency $f$. (g) Waterfall plot of \RHP\ as a function of \Bpara\ measured at different microwave frequency. Location of $r_1$ through $r_5$ are marked with vertical arrows.
}
\end{figure*}

Within the flat energy band of magic-angle graphene moir\'e systems, the influence of Coulomb interaction gives rise to prominent instabilities, which are commonly described by the spontaneous polarization in the internal flavor degrees of freedom of the moir\'e unit cell, given by spin, valley, and flatband degeneracy ~\cite{Park2021flavour,Zondiner2020cascade,Wong2020cascade}.
This process results in a reconstructed Fermi surface with well-defined isospin order. 
Since most emergent quantum phases, such as correlated insulators ~\cite{Lu2019SC,Cao2018a,Yankowitz2019SC}, superconductivity ~\cite{Cao2018b,Yankowitz2019SC,Lu2019SC} and topological ferromagnetism ~\cite{Sharpe2019,Serlin2019,Chen20201N2,Polshyn20201N2,Lin2021SOC}, are associated with different 
forms of flavor polarization, studying the process of flavor polarization, and the resulting isospin order, are essential to understanding the nature of electronic order in graphene moir\'e systems. 
Conventionally, experimental efforts rely on the evolution of the energy gap with in-plane magnetic field to determine the underlying isospin configuration. Given the requirement of a robust energy gap and well-defined thermal activation behavior, this method is only applicable to insulating states. Moreover, the multi-dimensional phase space of isospin order is defined by spin and valley isospin degrees of freedom. In the case of simultaneous presence or close competition of different symmetry-breaking orders, which is often the case in magic-angle graphene moir\'e systems, the $B$-dependence of the gap might not be enough to reveal the ground state order.
This lack of viable experimental methods contributes to the sizable gap in our understanding of the moir\'e flatband. For instance, intervalley Hund's interaction $J_H$ describes the coupling of electron spins across opposite valleys. The value and sign of $J_H$ are crucial for a wide variety of theoretical studies of superconductivity and correlated insulators in magic-angle graphene moir\'e systems ~\cite{Isobe2018tblg,Scheurer2020pairing,Khalaf2021C2T,Christos2020C2T,Kang2021cascades,Khalaf2020collective,Bernevig2021collective,Christos2021,Huang2021paramagnon}. However, even the sign of $J_H$ remains unknown up to now, owing to the scarcity of experimental constraints.

The ability to examine collective excitations could establish a new route to identify electronic orders within a moir\'e flatband, since the nature of these collective excitations is determined by the isospin order of the underlying ground state ~\cite{Khalaf2020collective,Bernevig2021collective,Huang2021paramagnon,Kumar2021collective,PhysRevB.100.035413}.  The existence of these collective modes has been hinted at by the observation of large electronic entropy within the moir\'e flatband ~\cite{Saito2021pomeranchuk,Rozen2021pomeranchuk,Liu2022DtTLG}. It is argued that the tendency to flavor polarize gives rise to fluctuating local flavor moments at high-temperature, which contributes to the large residual entropy and the associated Pomeranchuk-like transition. Furthermore, collective isospin fluctuations might also provide the pairing glue or at least crucially influence the form and strength of superconductivity.  
This highlights the importance of studying collective excitations in order to better understand the nature of the electronic orders in the moir\'e flatband.

In this work, we report the observation of collective excitations in magic-angle twisted bilayer graphene (MATBG) using the resistively-detected electron-spin resonance technique.
Collective excitations are generated when the energy of microwave photons matches the energy of the collective mode.
If the generation of the excitation gives rise to a change in the sample resistivity, it is detectable using standard DC transport techniques (see SI for more details regarding DC transport measurement).
The observation of electron-spin resonance establishes the first direct experimental identification of collective excitations in MATBG, which allows us to identify the nature of the spin coupling across opposite valleys and extract parameters that characterize spin properties.
We will focus on the density range of moir\'e filling $\nu$ with $2<|\nu|<3$, where multiple possible candidate orders \cite{Christos2020C2T} can capture the `Dirac revivals' \cite{Zondiner2020cascade,Wong2020cascade} at $|\nu|=2$.
By examining the microwave-induced resonance features, which are mostly unaltered in this density range, we extract important information on the underlying isospin order.

\begin{figure*}
\includegraphics[width=0.75\linewidth]{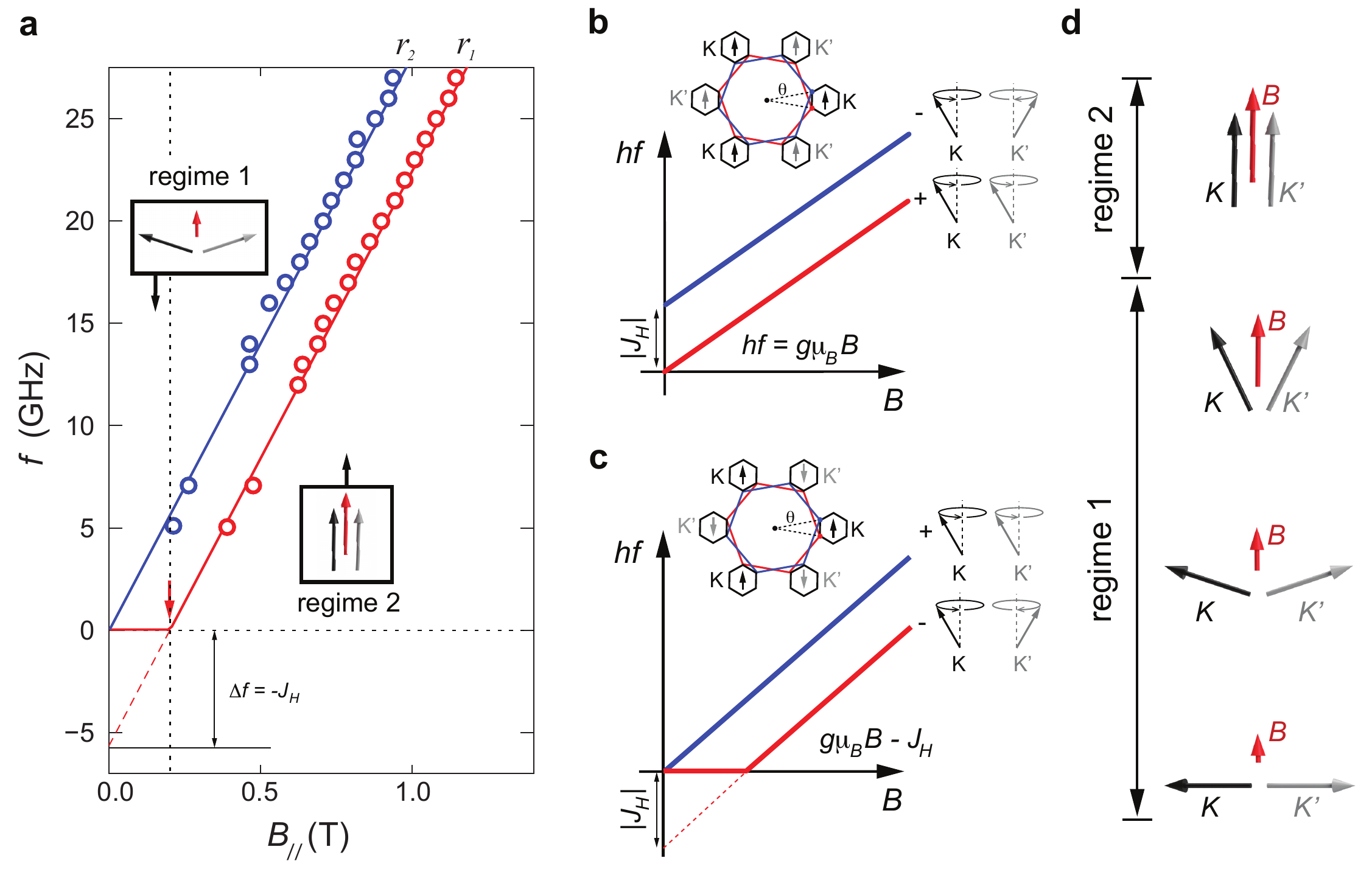}
\caption{\label{fig2} {\bf{Antiferromagnetic intervalley coupling and the value of $J_H$.}} (a) The location of $r_1$ and $r_2$ in the $f-B$ trajectories, fit by a spin magnon model where the primary mode extrapolates to a negative intercept $\Delta f = -J_H/h$, with intervalley Hund's coupling $J_H$. (b-c) A schematic of energy $hf$ versus $B$ trajectories for (b) ferromagnetic coupling  ($J_H<0$) in which spins are aligned in the two valleys (and along the external magnetic field) and (c) antiferromagnetic coupling ($J_H>0$); here, the anti-parallel spins develop an increasingly large ferromagnetic component along the field (regime 1) and perfectly align with it in the two valleys above a critical field (regime 2), as illustrated in (d). 
Insets of (b-c) depict schematics of the moir\'e Brillouin zone showing the nature of intervalley spin order.
}
\end{figure*}

The setup for the resonance measurement is illustrated in Fig.~\ref{fig1}a-b.
A coaxial line is placed above and perpendicular to the MATBG sample shaped into a Hall bar geometry. As a microwave signal is applied to the coaxial line between $1$ and $28$ GHz with source powers between $-120$ and $21$ dBm, it irradiates the device with microwave photons.
Fig.~\ref{fig1}c-g show transport response across a MATBG sample in the presence of an in-plane magnetic field with varying power and frequency in the microwave radiation at moir\'e filling $\nu=2.8$, which falls within the density range of $2 < |\nu| < 3$. Microwave radiation induces prominent changes in the longitudinal resistance of the sample, \DRxx. 
At a low microwave power, $P = -31$ dBm at $f= 5$ GHz, \DRxx\ exhibits a single dip, which is marked by the vertical dashed line in Fig.~\ref{fig1}c. The location of this dip follows a linear trajectory in the $f-B$ map (Fig.~\ref{fig1}d). At higher microwave power, the dip in \DRxx\ becomes more pronounced (upper panel of Fig.~\ref{fig1}e). At the same time, this transport response widens to occupy a larger area in the $f-B$ map, as shown in Fig.~\ref{fig1}f. The fact that increasing microwave power produces more prominent changes in sample resistance suggests that these features originate from the coupling between microwave photons and electrons in the moir\'e flatband. The application of a high-pass filter (HPF) eliminates the slow-varying background in \DRxx\ as a function of an in-plane magnetic field, which yields \RHP\ (see Fig.~\ref{figHPF} for comparison between HPF and derivative $dR/dB$).
While \RHP\ exhibits similar behavior as \DRxx\ at low microwave power, it reveals five separate features at high power, which are manifested as sharp minima in \RHP. The location of these features are marked by vertical dashed lines and white arrows in Fig.~\ref{fig1}c-g. For simplicity, we will refer to these features as resonance modes $r_1$ through $r_5$. 

Two observations testify that $r_1$ is the fundamental resonance mode.  First, the dependence of these resonances on varying microwave power shows that $r_1$ emerges at the lowest microwave power over all frequencies (see Fig.~\ref{fig:SI_power}). Second, the microwave-power-dependence shows that the resonance mode with a lower excitation energy, defined by its location in the $f-B$ map at a lower frequency, is more robust compared to one with higher energy, with $r_1$ (\red{$r_4$}) being the most (least) robust  (see Fig.~\ref{fig1}c-e). One of the main focuses of this work is to understand the origin of $r_1$. We will also examine the condition required for detecting electron spin resonance using resistive methods in encapsulated graphene samples. Furthermore, we offer experimental characterization and theoretical interpretation for all other resonance modes. 
 
The energy of the microwave photon provides an important clue regarding the origin of the observed resonance modes. A $10$ GHz microwave photon corresponds to an energy of $0.04$ meV. The observation of the resonance response indicates that electrons must be able to absorb microwave photons and transition into an excited state, producing a change in sample resistivity at the same time.
Owing to the strong Coulomb correlation in a moir\'e flatband, the energy gap associated with flavor polarization and Fermi surface reconstruction is on the order of several meV ~\cite{Cao2018a,Yankowitz2019SC,Liu2021DtBLG}. As a result, particle-hole excitations across these correlation-driven gaps require an energy that far exceeds the photon energy. As such, the excited states corresponding to $r_1$ through $r_5$ must be associated with low-energy collective excitations that are below the continuum of interband particle-hole excitations. Since an in-plane magnetic field is mostly decoupled from valley and sublattice degrees of freedom ~\cite{Sharpe2021}, the collective excitation must emerge from the spin channel. 

Fig.~\ref{fig1}g shows the waterfall plot of \RHP\ versus an in-plane magnetic field \Bpara\ measured at different microwave frequency $f$. The position of $r_1$ through $r_5$ are identified as sharp minima. The evolution of $r_1$ through $r_5$ with varying microwave frequency $f$ and magnetic field $B$ follows well-defined linear trajectories across a wide frequency range, $5 < f < 30$ GHz. 
We will first examine the location $r_1$, which are extracted from the waterfall plot in Fig.~\ref{fig1}g and marked with red circles in Fig.~\ref{fig2}a. The observed linear trajectory exhibits a slope that corresponds to an electron $g$-factor of $2$. This is strong indication that the transport response originates from an electron spin resonance. Most remarkably, the linear trajectory of $r_1$ extrapolates to a negative intercept with the frequency axis at $B=0$, which provides the finger print to identify the nature of this collective excitation.  

Given the phase space defined by spin, valley isospin and sublattice, it is recognized that the `Dirac revival' at half-filling could give rise to $15$ possible order parameters \cite{Christos2020C2T}.
Out of these options, the linear slope of $g=2$ is most naturally explained by two candidate orders---a state with parallel (ferromagnetic) spin polarization in the two valleys and an antiferromagnetic state where the spins are anti-aligned (see SI for more detailed discussion). 
The nature of the valley-dependent spin-configuration is determined by the intervalley Hund's interaction $J_H$, with $J_H<0$ (expected for Coulomb interactions \cite{Chatterjee2020}) and $J_H>0$, corresponding to parallel and anti-parallel spins, respectively.

In the presence of a ferromagnetic Hund's coupling with $J_H < 0$,  there will be a single Goldstone mode for vanishing magnetic field $\vec{B}=0$, as a consequence of the spontaneously broken spin-rotation symmetry. While gapless at $\vec{B}=0$, this mode will exhibit a finite gap $\Delta=g\mu_B |\vec{B}|$ when $\vec{B}\neq 0$, determined by the spin Zeeman energy. The resonance behavior of this Goldstone mode is shown as the red solid line in Fig.~\ref{fig2}b. 
In the case of $J_H <0$, the $|\vec{B}|$ dependence of all microwave resonance frequencies will be linear and, for those associated with single magnon processes, of slope $g\mu_B/h$. The intercept will be zero for the resonance with the lowest energy, which appears at the lowest microwave frequency. This behavior is not consistent with experiment where the lowest and most dominant resonance mode exhibits a negative intercept. 

Let us therefore consider the case $J_H>0$, where the spins in opposite valleys are anti-parallel at $\vec{B}=0$ (inset in Fig.~\ref{fig2}c).  When we turn on the external magnetic field $\vec{B}\neq 0$, anti-parallel spins are canted with their ferromagnetic (anti-ferromagnetic) components along (perpendicular) to $\vec{B}$, see regime 1 in Fig.~\ref{fig2}d. When $g\mu_B|\vec{B}| \geq J_H$ (regime 2), the anti-ferromagnetic component vanishes and the spins are fully aligned with the magnetic field.
In the anti-ferromagnetic case, the resonance behavior of lowest energy excitation is shown as the red solid line in Fig.~\ref{fig2}c.
It remains gapless for $\vec{B}\neq 0$ in regime 1, which derives from breaking the single continuous symmetry given by spin rotations along the magnetic field axis \cite{CountingGoldstones}. In regime 2, however, it develops a non-zero gap given by $\Delta = g\mu_B |\vec{B}| - J_H$, which extrapolates to $B=0$ with a negative intercept of $\Delta f =-J_H$. This is in excellent agreement with the observed behavior of $r_1$, providing strong indication for $J_H>0$.  
Most remarkably, the measured intercept from mode $r_1$, which corresponds to the Goldstone mode $-$ in regime 1, allows us to extract the value of $J_H$. A negative intercept of $\Delta f = -5.6$ GHz (Fig.~\ref{fig2}b) yields an intervalley Hund's coupling, $J_H = 0.023$ meV, which is compatible with numerical estimates of $J_H = 0.075$ meV ~\cite{Chatterjee2020}.
 
Along the same vein, $r_2$ can be understood by simply considering the valley degrees of freedom.
Due to the valleys, the spin in the two valleys can be either in-phase for the resonance mode, or out-of-phase. For simplicity, we will refer to the in-phase and out-of-phase modes as $p=+$ and $-$, respectively. In the case of $J_H < 0$, mode $+$ ($-$) has the lower (higher) energy, due to ferromagnetic coupling. The energy gap of the $-$ mode is given by $g\mu_B |\vec{B}| + |J_H|$, which remains finite at $|\vec{B}|=0$ (blue solid line in Fig.~\ref{fig2}b). 
The energy sequence of the first two modes are switched in the case of $J_H>0$. The lowest energy mode is $-$, which is favored by the antiferromagnetic coupling. The energy gap of mode $+$ is given by the Zeeman energy. It corresponds to a straight line with slope $g\mu_B/h$ and vanishing intercept (blue solid line in Fig.~\ref{fig2}c), which is consistent with the trajectory displayed by the $r_2$ mode. Compared to $r_1$, $r_2$ requires a higher microwave power to be detectable (Fig.~\ref{fig:SI_power}). This is consistent with the overall hierarchy among all resonance signals.

\begin{figure*}

\includegraphics[width=1\linewidth]{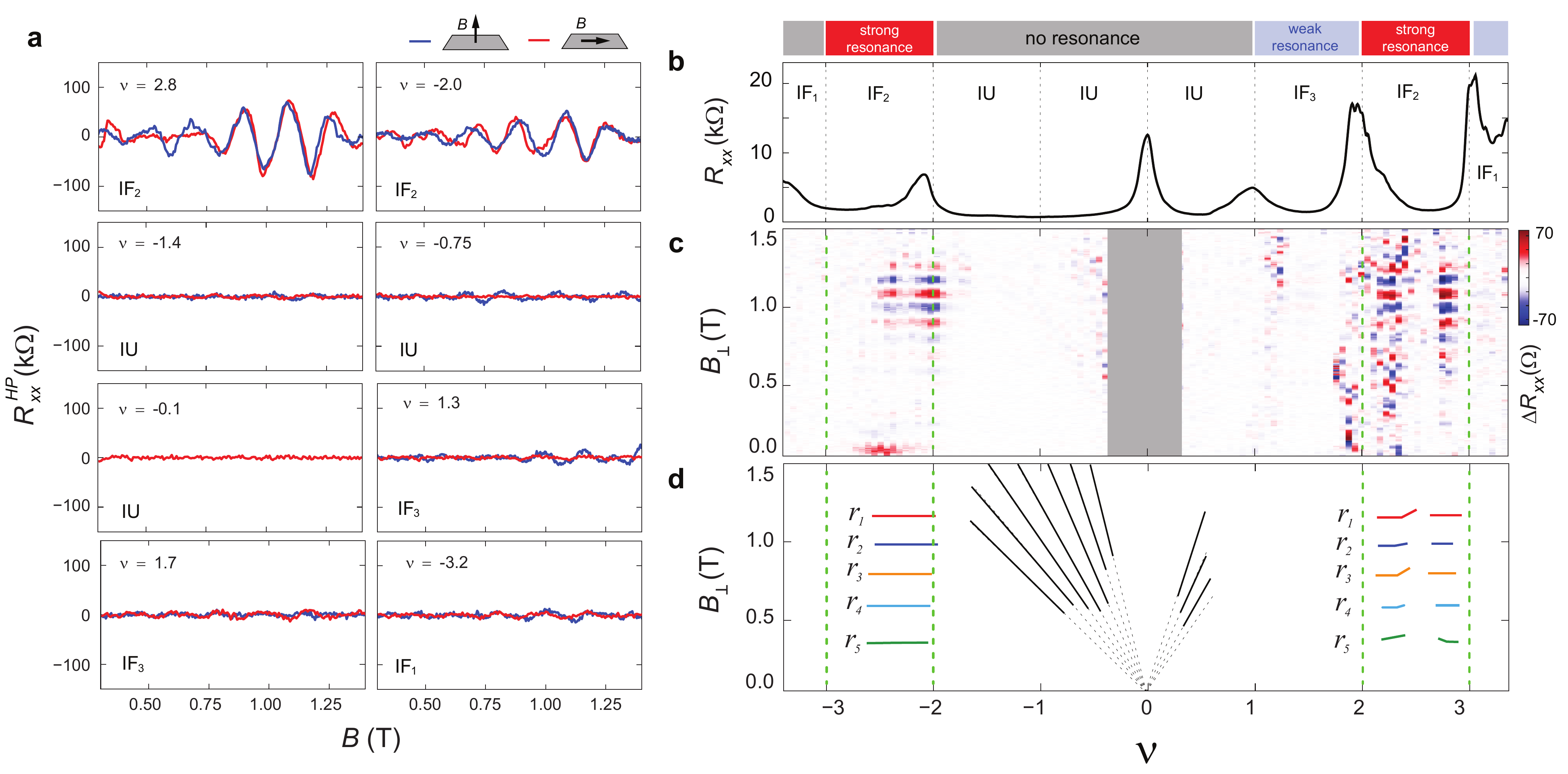}
\caption{\label{figN} 
{\bf{Density dependence and magnetic anisotropy.}} (a) \RHP\ as a function of \Bperp\ and \Bpara, measured at a range of different moir\'e fillings in the presence of microwave radiation with $f = 27$ GHz. The sample is highly resistive at the CNP ($\nu=-0.1$) in the presence of an out-of-plane $B$-field. Therefore, only the dependence on \Bpara\ is plotted. The slight oscillation in the \Bperp-dependence at $\nu=-1.4$ and $-0.75$ derives from quantum oscillation emanating from the CNP (see Fig.~\ref{fig:LL}c).  (b) \Rxx\ as a function of filling measured at $T = 20$ mK and $B = 10$ mT. Resistance peaks in \Rxx\ separate symmetry-breaking isospin ferromagnets (IF$_2$ and IF$_3$) and isospin-unpolarized states (IU). The density range of observed microwave-induced resonance is marked on the top axis. (c)\RHP\ as a function of filling fraction $\nu$ and \Bperp\ measured at microwave frequency $f=27$ GHz. (d) The location of resonance $r_1$ through $r_5$ marked in a schematic $\nu-B$ map.  Solid colored lines denote the location of resonance modes in the density range of $2 < |\nu| < 3$, which is the focus of this study, whereas dashed colored lines are resonance modes outside of this density range. Black lines mark the location of quantum Hall effect states observed near the CNP. }
\end{figure*}

In the following, we show the dependence of resonance response on moir\'e filling, which reveals a direct link between the flat moir\'e band and the observed resonance behavior ~\cite{Park2021flavour,Zondiner2020cascade,Saito2021pomeranchuk}. 
Fig.~\ref{figN} plots \RHP\ at $f = 27$ GHz as a function of both in-plane and out-of-plane magnetic field, \Bpara\ and \Bperp, measured at different moir\'e fillings. Prominent resonance response is observed near half-filling of both electron and hole-doped moir\'e bands, at $\nu= 2.8$ and $-2$. At these moir\'e fillings, the resonance signal remains the same for both out-of-plane and in-plane magnetic field, which is consistent with excitations from the spin channel (see Fig.~\ref{fig:Bdirection}). Notably, the presence of strong spin-orbit coupling will lead to resonance behaviors which generically depend on the $B$-field orientation. The fact that the resonance frequencies are unaltered when the magnetic field is rotated in and out of the plane of the system indicates that the influence of spin-orbit coupling is not crucial for the spectrum of the excitations giving rise to the microwave resonance features (see Fig.~\ref{figN}a and Fig.~\ref{fig:Bdirection}). We note, however, that despite the weak impact of spin-orbit coupling onto the spectrum of these modes, symmetry reduction due to the nearby WSe$_2$ layer could be crucial for an efficient coupling of microwaves and the collective modes, and their impact on transport. 

Most interestingly, transport measurement shows no resonance response near the neutrality point (CNP). In this regime, slight variation in \RHP\ develops with increasing out-of-plane magnetic field, which corresponds to quantum oscillation emanating from the CNP. Similarly, resonance response is absent, or extremely weak, around one-quarter and three-quarters fillings. 
To gain more insights into the density dependence of resonance behavior, 
Fig.~\ref{figN}b-c compare the resonance response, shown as horizontal features with blue and red colors in Fig.~\ref{figN}c, with the dependence of longitudinal resistance \Rxx\ on moir\'e filling $\nu$ across the moir\'e flatband. As shown in Fig.~\ref{figN}b, the emergence of resistance peaks near integer moir\'e fillings of $\nu=-2$, $+1$, and $+2$ is in excellent agreement with previous experiments on MATBG ~\cite{Saito2021pomeranchuk,Park2021flavour,Lin2021SOC,Zondiner2020cascade,Rozen2021pomeranchuk}. These resistance peaks, along with resets in Hall density (see Fig.~\ref{fig:Hall}) define the boundaries of different isospin orders resulting from Dirac revival. The isospin order of the underlying Fermi surface away from integer filling can be identified based on the main sequence of quantum oscillations (see Fig.~\ref{fig:LLhighB} and Fig.~\ref{fig:LLlowB}). Based on these characterization, we will label different regimes of the moir\'e band fillings according to the underlying isospin order, such as isospin-unpolarized (IU) and isospin ferromagnet with N-fold degeneracy, IF$_N$ (Fig.~\ref{figN}c). Here $N$ takes the value of $1$, $2$ and $3$, which corresponds to the degeneracy in the quantum oscillation. 
Strong resonance responses in Fig.~\ref{figN}c show excellent agreement with the density range of IF$_2$ on both electron and hole-doping side of the moir\'e band. The boundaries of IF$_2$ are marked with the green vertical dashed lines in Fig.~\ref{figN}c-d. 
Saturating the color scale of Fig.~\ref{figN}c reveals extremely weak resonance response in the IF$_3$ regime (see Fig.~\ref{fig:LL}). However, there is no indication of any resonance signal throughout the density range of IU (Fig.~\ref{figN}c and Fig.~\ref{fig:LL}). In this regime, the only detectable transport features are quantum oscillations emanating from the CNP. 
The robust resonance at half-filling, contrasting with the lack of resonance response near the CNP, places an important constraint on the possible origin of the resonance response. 
The onset of resonance at $\nu=\pm2$ is a clear evidence of an origin in strong correlations within the moir\'e flat band. 
A natural explanation is that these are collective excitations associated with textures in the flavor degrees of freedom, such as magnon modes shown in Fig.~\ref{fig2}c.

In the density range identified as IF$_3$ or IF$_1$, microwave-induced resonance response is mostly absent, despite the presence of ``Dirac revival''. The lack of resonance in these regimes may be the result of a few different factors. First, the magnitude of the resonance behavior could depend on the strength of isospin flavor polarization. In MATBG, half-filling at $\nu \pm 2$ often host the most prominent correlation-driven states with the most robust energy gap, whereas the energy gaps at quarter and three-quarter fillings are only partially developed. Second, the different isospin order of IF$_3$ and IF$_1$ may give rise to a distinct response to microwave radiation.  For instance, the resonance response at $\nu=1.3$ exhibits strong dependence on $B$-field orientation: although weak resonance features are detectable with an out-of-plane $B$-field, there is no resonance signal in the presence of an in-plane $B$-field (see Fig.~\ref{figN}a). 
Given the distinct behavior, the mechanism underlying the extremely weak signal at $\nu=+1.2$ likely differs from $\nu=\pm2$. Since the latter is the main focus of this work, we will leave further investigation of the former response for future efforts. 

In the following, we will compare MATBG with other graphene  monolayer and bilayer samples, including twisted bilayer graphene, where the energy band is highly dispersive. In these samples,  flatband physics such as Dirac revival is absent and the underlying Fermi surface is known to be isospin unpolarized, making them directly comparable with the IU regime of MATBG. These graphene samples all feature doubly-encapsulated geometry with hexagonal boron nitride (hBN) and graphite encapsulation to minimize the influence of charge fluctuation and outside impurities on the resonance measurement (see SI for more detailed sample characterization). Despite the excellent sample quality, spin resonance measurements on these graphene monolayer and bilayer samples, using the same setup as shown in Fig.~\ref{fig1}a, show no indication of any microwave-induced resonance signal over a large range of microwave frequency, power, and charge carrier density across both electron and hole-type carrier polarities (see SI, Fig.~\ref{fig:MWfans}, Fig.~\ref{fig:MWdif}, Fig.~\ref{fig:monoline}. Fig.~\ref{fig:MLG2}, Fig.~\ref{figMWMSU} and Fig.~\ref{figBernal} for more detailed discussions). 
These observations offer further confirmation that the strong resonance response in MATBG is intrinsic to the flat moir\'e energy band and directly associated with the isospin order underlying the ``Dirac revival'' at $\nu=\pm2$.  
The lack of resonance response in high quality graphene monolayer and bilayer samples is in stark contrast with previous spin-resonance experiments \cite{Mani2012ESR,Sichau2019ESR,Blick2020} on unencapsulated monolayer graphene (MLG) samples prepared with the chemical vapor deposition method, which have known material issues ~\cite{Mani2012ESR,Sichau2019ESR}.
It has been well-documented in the literature that transport response across graphene is highly susceptible to the influence of outside disorder and impurity. As such, encapsulation with hBN and graphite is essential to probing intrinsic behaviors of graphene  ~\cite{Lei.13,Zibrov2017,Li.17b,Zeng2019Corbino,Polshyn2018Corbino}. 

Taken together, our understanding for the origin of $r_1$, which is associated with the intervalley exchange interaction $J_H$ in the IF$_2$ regime, is supported by a series of prominent experimental characteristics: (i) $r_1$ is the fundamental resonance mode that is detectable at the lowest microwave power; (ii) the resonance behavior of $r_1$ tracks a linear trajectory in the $f-B$ map which extrapolates to a negative intercept at $B=0$; (iii) the behavior of $r_1$ remains the same with in-plane and out-of-plane magnetic field; (iv) prominent resonance response is only observed in the density regime that corresponds to an isospin ferromagnet with 2-fold degeneracy (IF$_2$), whereas no resonance is observed in the isospin unpolarized regime (IU); (v) using the same experimental setup, we observe no indication of resonance response in graphene monolayer and bilayer samples without a flat moir\'e band.

\begin{figure*}
\includegraphics[width=0.7\linewidth]{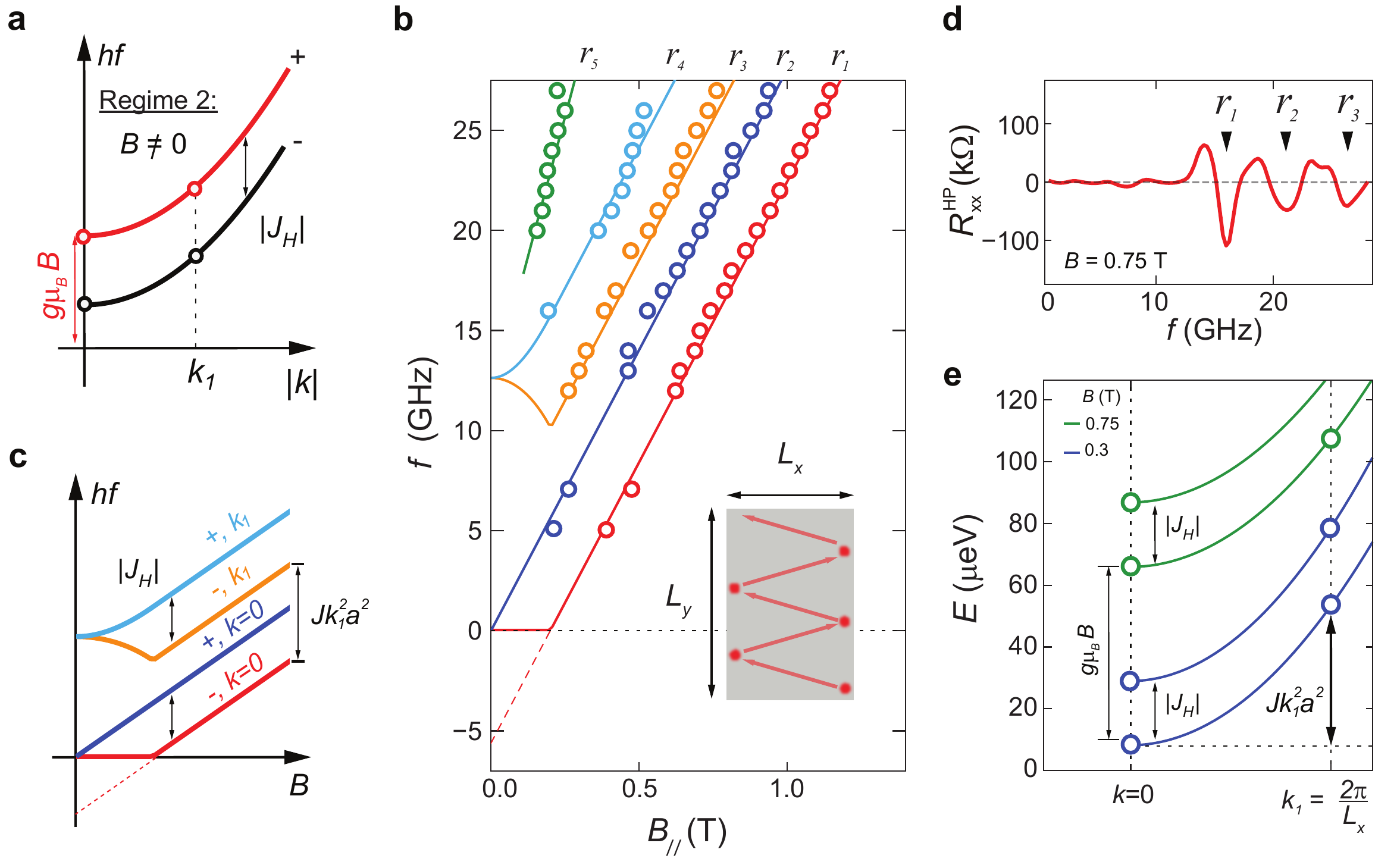}
\caption{\label{fig3} {\bf{Spin stiffness and dispersion.}} (a) Schematic of the magnon dispersion in regime 2, characterized by field $B$ as well as $J_H$.  (b) The location of $r_1$ through $r_5$, marked with red, blue, orange, cyan and green circles, in the $f-B$ map. The solid lines correspond to the theoretical fit for the resonance modes. (c) Schematic of magnon spectrum resulting from an antiferromagnetic intervalley coupling. Red and blue lines correspond to the $k=0$ modes, whereas orange and cyan lines denote geometric resonance modes with $k \neq 0$. The offset between the $k=0$ and $k_1$ modes is characterized by $J k_1^2 a^2$ where $J$ is a measure of spin stiffness and $a$ is the moir\'e lattice constant. (d) Vertical linecut from the $f-B$ resonances at $B = 0.75$~T, where minima in \RHP\ are the locations of the first three modes $r_1$ through $r_3$. (e) The location of the resonance features from \RHP $-f$ linecuts such as in (d), and where modes $r_1, r_2$ correspond to $k=0$ and $r_3, r_4$ to $k_1 \neq 0$. 
}
\end{figure*}

Having established a theoretical model for understanding the $r_1$ and $r_2$ modes, we are now in position to investigate higher order resonance modes $r_3$ through $r_5$. While we cannot rigorously exclude that some of these modes have an origin distinct from $r_1$ and $r_2$, we next show that they can be naturally captured by the magnon picture outlined above, providing for further support for the latter.
Magnon modes are universally associated with an energy spectrum (Fig.~\ref{fig3}a and Fig.~\ref{fig:SI_theory}).
Both $r_1$ and $r_2$ correspond to the zero momentum ($\vec{k}=0$) limit of this magnon spectrum. 
At small $\vec{k}$, the spectrum is quadratic, $E_{\vec{k}} \sim J \vec{k}^2 a^2$, in regime 2, where $a$ is the moir\'e lattice constant and $J$ provides a measure of the spin stiffness (see SI for more detailed  definition of $J$). 
The finite size of the sample allows spin resonance to occur at a set of discrete values of the momentum vector. Taking the system geometry to be rectangular of size $L_x \times L_y$, these discrete momentum values are given by $\vec{k}=2\pi (n/L_x,m/L_y)^T$, $n,m\in\mathbb{Z}$. For a sample that is $2$  $\mu$m in dimension, the first geometric resonance mode occurs at $k_1 = 3.14$ $\mu$m$^{-1}$. In the $f-B$ map, this geometric resonance mode gives rise to two extra resonance modes, $+,k_1$ and $-,k_1$, as shown in Fig.~\ref{fig3}c, which correspond to $r_3$ and $r_4$ in our observation (Fig.~\ref{fig3}b).
The offset between $+,k_1$ and $-,k_1$, given by $|J_H|$, is expected to be the same as the offset between $+,k=0$ and $-,k=0$. This is in excellent agreement with the observed behavior of $r_3$ and $r_4$ in Fig.~\ref{fig3}b.
Notably, the location of the higher-frequency modes $r_3$ and $r_4$ can shed light on the magnon dispersion, offering experimental constraints on parameters such as the spin stiffness and the propagation velocity of magnons. Fig. \ref{fig3}d shows a linecut of \RHP\ as a function of frequency, or energy, taken from the $f-B$ map at $B = 0.75$ T. The location of resonance features are indicated by sharp minima in \RHP. Plotting energy versus momentum corresponding to either $k=0$ ($r_1, r_2$) or $k \neq 0$ ($r_3, r_4$) as in Fig. \ref{fig3}e yields a fit of the energy spectrum (solid lines), with 
the only fitting parameter beyond $J_H$ being spin stiffness $J$. Given the momentum of the first geometric resonance mode at $k_1 = 3.14$ $\mu$m$^{-1}$, we extract the value of spin stiffness to be $J = 23$ meV.
From these fitted values for $J_H$ and $J$, we can also extract the magnon velocity in the low-field regime (regime 1), given by $v_B = \sqrt{J_H J a^2}\ |\cos \theta_B|$, where $\theta_B$ is the canting angle (see SI for more details). We find $v_B$ to be of the order of $10$ km/s at small fields.

While the simple model of magnon geometric resonance provides an excellent explanation for the observed resonance of $r_3$ and $r_4$, our Hall-bar-shaped sample is not the ideal geometry for detecting geometric resonance, since it lacks a well-defined boundary that could form a cavity. This likely accounts for the main source of error in estimating the spin stiffness $J$.  
Nevertheless, the extracted value for $J$ and $v_B$ have the correct order of magnitude when compared with previous experimental characterizations based on the electronic entropy measurement at high temperature ~\cite{Saito2021pomeranchuk,Liu2022DtTLG} 
and spin wave transport in a Fabry-P\'erot cavity ~\cite{Fu2021magnon}.
Furthermore, $r_5$ can be understood by considering a two-magnon mode, given its slope of $g=4$ (see SI for more details) ~\cite{Chinn1971twomagnon,Davies1972twomagnon}.

In summary, our findings imply an anti-ferromagnetic alignment of the spins in opposite values in the doping range $2\leq |\nu| \leq 3$ and provide the first experimental determination of the sign and value of the intervalley Hund’s interaction $J_H = 0.023$ meV. This is a crucial parameter which also determines the spin structure of the superconducting order parameter: for the anti-ferromagnetic sign we extract, the order parameter will be either spin singlet or a singlet-triplet admixed state \cite{Scheurer2020pairing}; the latter admixed state must be realized, if superconductivity coexists microscopically with the isospin order we identified. We note that this conclusion is consistent with a recent  analysis \cite{LakePairingPhen} of the body of other experiments on twisted bi- and trilayer graphene.
Moreover, the ability of our microwave study to probe $J_H$ could shed light onto the mechanism underlying the intervalley exchange interaction. It has been proposed that Coulomb interaction leads to a ferromagnetic coupling and a negative $J_H$, whereas the contribution from the electron-phonon interaction favors an anti-ferromagnetic coupling and a positive $J_H$ ~\cite{Chatterjee2020}. 
While our results appear to indicate that the electron-phonon coupling plays a more dominant role, combining microwave measurements with Coulomb screening in future experiments would allow us to untangle the role of Coulomb interaction and electron-phonon coupling in determining $J_H$. Finally, our microwave resonance probe of collective modes would also enable us to identify exotic emergent symmetries in other twisted graphene superlattices \cite{2022arXiv220405317W}.

\section*{Acknowledgments}

A.M and J.I.A.L. thank R. Cong for critical review of the manuscript.
E.M. acknowledges funding from the National Defense Science and Engineering Graduate (NDSEG) Fellowship. 
J.-X.L. and J.I.A.L. acknowledge funding from NSF DMR-2143384. Device fabrication was performed in the Institute for Molecular and Nanoscale Innovation at Brown University. J.P. and L.Z. acknowledge support from the Cowen Family Endowment at MSU.
K.W. and T.T. acknowledge support from the EMEXT Element Strategy Initiative to Form Core Research Center, Grant Number JPMXP0112101001 and the CREST(JPMJCR15F3), JST.
Sandia National Laboratories is a multi-mission laboratory managed and operated by National Technology and Engineering Solutions of Sandia, LLC, a wholly owned subsidiary of Honeywell International, Inc., for the DOE’s National Nuclear Security Administration under contract DE-NA0003525.  This work was funded, in part, by the Laboratory Directed Research and Development Program and performed, in part, at the Center for Integrated Nanotechnologies, an Office of Science User Facility operated for the U.S. Department of Energy (DOE) Office of Science. This paper describes objective technical results and analysis. Any subjective views or opinions that might be expressed in the paper do not necessarily represent the views of the U.S. Department of Energy or the United States Government.

\bibliography{Li_ref}%

\newpage

\newpage
\clearpage

\pagebreak
\begin{widetext}
\section{Supplementary Materials}

\begin{center}
\textbf{\large Low-energy collective modes and antiferromagnetic fluctuation in magic-angle twisted bilayer graphene}\\
\vspace{10pt}
Erin Morissette,
Jiang-Xiazi Lin, 
Dihao Sun, Liangji Zhang,
  Song Liu, Daniel Rhodes, Kenji Watanabe, Takashi Taniguchi, James Hone, Johannes Pollanen, Mathias S. Scheurer,
Michael Lilly, Andrew Mounce, J.I.A. Li$^{\dag}$\\ 
\vspace{10pt}
$^{\dag}$ Corresponding author. Email: amounce@sandia.gov, jia$\_$li@brown.edu
\end{center}

\noindent\textbf{This PDF file includes:}

\noindent{Supplementary Text}

\noindent{Materials and Methods}

\noindent{Figs. S1 to S18}

\renewcommand{\thefigure}{S\arabic{figure}}
\setcounter{figure}{0}
\setcounter{equation}{0}

\newpage

\section{Supplementary text}

In this work, we have studied the dependence of the fundamental resonance mode $r_1$ on microwave frequency,  microwave power, in-plane and out-of-plane magnetic field, and moir\'e filling. The theoretical analysis of the underlying mechanism is based on the combination of all measurements, including, but not limited to, the location of the resonance mode in the $f-B$ map. At the same time, we have performed a microwave resonance measurements on a series of monolayer and bilayer samples without a flat energy band, where we do not observe any resonance modes. Here we summarize all the experimental findings.

\begin{itemize}

\item{We observe a hierarchical behavior amongst resonance modes, where a mode with lower excitation energy is more robust compared to one with higher excitation energy. This hierarchy is demonstrated by two separate observations: (i) a resonance mode with lower excitation energy corresponds to more prominent transport response; (ii) a resonance mode with lower excitation energy persists to lower microwave power. Combined, our findings indicate that $r_1$ is the fundamental fundamental resonance mode. }

\item{The observed fundamental mode $r_1$ tracks a trajectory in the $f-B$ that exhibits a negative intercept. This is distinct from a potential ferromagnetic resonance (or paramagnetic resonance), where the most robust resonance response has a diminishing intercept at $B = 0$. }

\item{Prominent microwave resonance response is only observed in the density regime that corresponds to an isospin ferromagnet with 2-fold degeneracy (IF$_2$). No resonance is observed in the isospin unpolarized regime (IU). This suggests that the resonance behavior is directly linked to the Dirac revival at $\nu=\pm2$ and the resulting isospin order.}

\item{In the IF$_2$ regime, resonance response is shown to be insensitive of the orientation of external magnetic field. $r_1$ through $r_5$ remain the same with an out-of-plane and in-plane magnetic field. This provides a strong indication that the influence of spin-orbit coupling is negligible. }

\item{Using the same microwave setup, we have performed resonance measurements on a series of graphene monolayer and bilayer samples, where the energy band is dispersive. In the absence of flatband physics such as Dirac revival, we observe no indication of microwave-induced resonance over a large range of carrier density, microwave frequency and microwave power.}

\end{itemize}

In the following, we provide more details related to these measurements that will supplement the discussions in the main text.

\subsection{Microwave power dependence, hierarchical behavior and the location of the resonance mode}

\begin{figure*}[t]
\includegraphics[width=0.85 \linewidth]{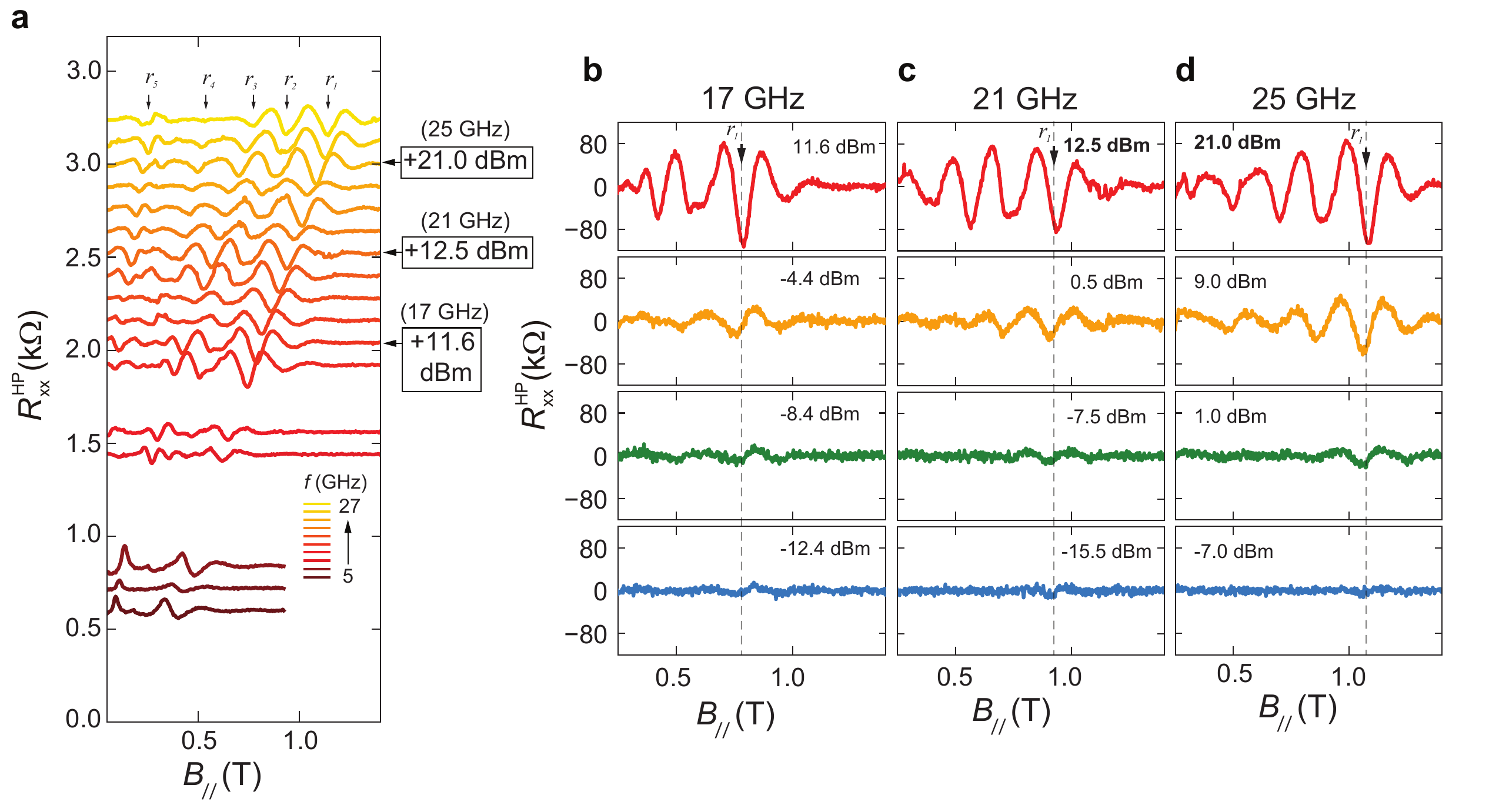}
\caption{\label{fig:SI_power} {\bf{Dependence of resonance modes on microwave power.}} (a) Waterfall plot of \RHP as a function of \Bpara, in which each microwave frequency line is calibrated to a qualitatively equivalent microwave power. 
Three frequency lines and their respective power spectra are illustrated in (b) for 17 GHz where the microwave power ranges from $-12.4$ to $11.6$ dBm, (c) $21$ GHz from $-15.5$ to $12.5$ dBm, and (d) $25$ GHz from $-7$ to $21$ dBm.
}
\end{figure*}

\begin{figure*}[h]
\includegraphics[width=0.36\linewidth]{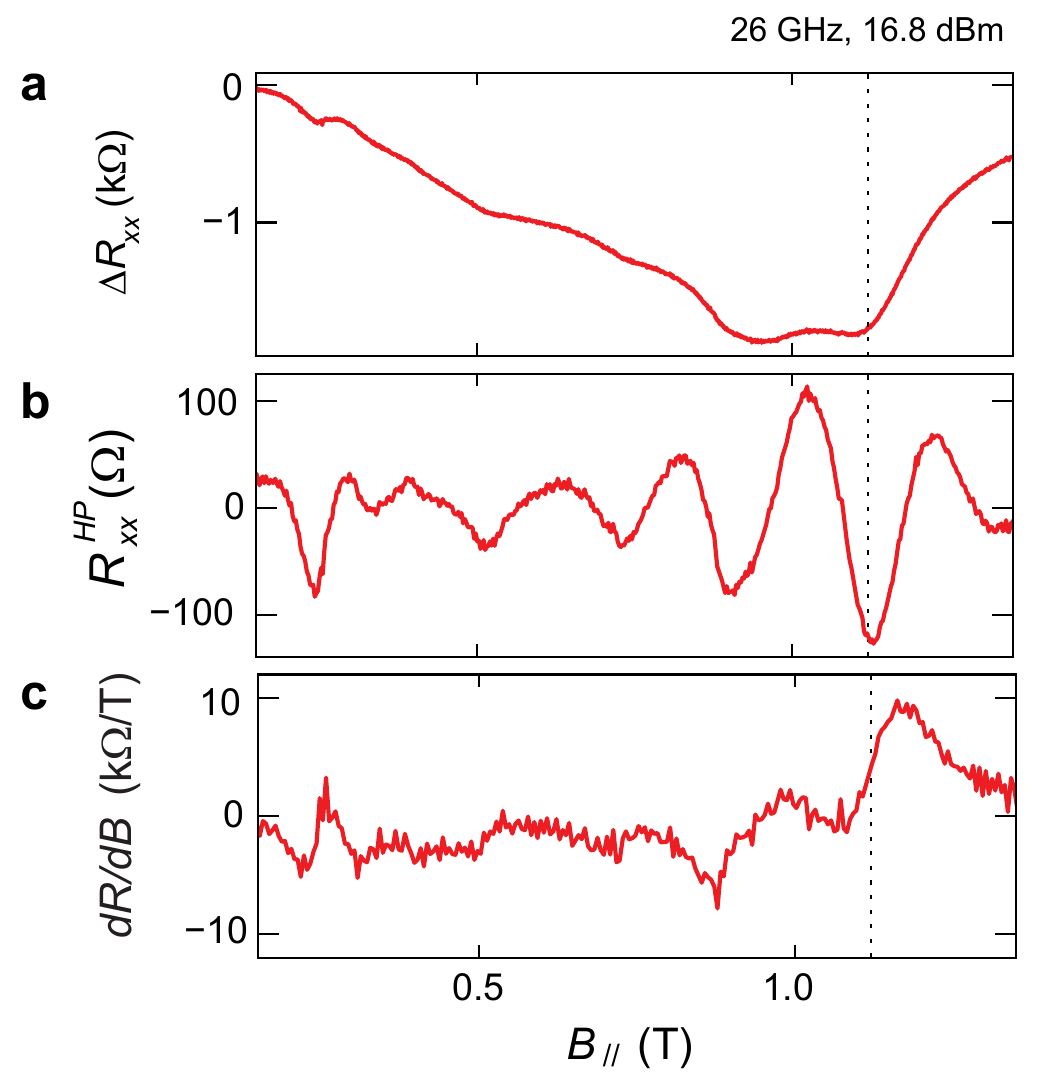}
\caption{\label{figHPF} {\bf{High-pass filter and $\boldsymbol{dR/dB}$.}} (a) The change in longitudinal resistance, \DRxx, with in-plane field for 26 GHz at a high power. (b) The derivative $dR/dB$, and (c) the high-pass filtered result \RHP\ of the signal shown in panel (a). The rightmost dip in \DRxx\ ($r_1$) is denoted by a dashed line. The high-pass filtered response maintains the Gaussian-like shape of the original feature as well as the $B$-position of the dip. On the other hand, in the derivative channel, a Gaussian-shaped resonance splits into a peak and a dip. The location of the resonance mode corresponds to the inflection point in the derivative channel. Taking the derivative could introduce error in identifying the location of the resonance mode in the $f-B$ map, whereas the high-pass filter approach does not introduce this type of error. 
}
\end{figure*}

As noted in the main text, $r_1$ through $r_4$ demonstrate a well-defined hierarchical behavior, with $r_1$ ($r_4$) being the most (least) prominent resonance mode in our measurement. 
The hierarchy is demonstrated by two separate measurements, as shown in Fig.~\ref{fig:SI_power}b-d. First, for all microwave frequency, $r_1$ corresponds to the most prominent response in transport measurement. In fact, the magnitude of transport response decreases gradually from $r_1$ to $r_4$. Second, the stability of the resonance mode is reflected by the lowest microwave power where the mode is detectable. For instance, at $f=25$ GHz, $r_1$ is the only mode present at $-7$ dBm. Upon increasing microwave power to $-1$ dBM, $r_2$ becomes visible. This provides further indication that $r_1$ is a more robust resonance mode compared to $r_2$. Fig.~\ref{fig:SI_power}b-d shows that the same hierarchical behavior is observed at all frequencies.  $r_2$ through $r_4$ becomes detectable with increasing power.  
This hierarchical behavior suggests that it is more likely for electrons to absorb a photon and  transition into an excited state with a lower excitation energy, such as $r_1$, compared to one with a higher excitation energy, such as $r_2$. This hierarchy indicates that the mechanism underlying the $r_1$ mode plays the dominating role in the generation and detection of microwave resonance. Combined with the dependence on the $B$-field orientation (see Fig.~\ref{fig:Bdirection}) and moir\'e filling (Fig.~\ref{figN}a-c), a resonance with antiferromagnetic intervalley coupling provides the most natural explanation. 
This is distinct from the scenario where ferromagnetic, or paramagnetic, resonance plays the dominating role. In this case, the most prominent resonance mode is expected to have diminishing intercept, which corresponds to the trajectory of $r_2$ in the $f-B$ map.

Fig.~\ref{figHPF} compares the resonance response in the longitudinal resistance channel with the derivative $dR/dB$ and the high-pass-filtered response \RHP. The high-pass filtered response maintains the Gaussian-like shape of the original feature as well as the $B$-position of the dip. This is in stark contrast with the response in the derivative channel, where a Gaussian-shaped resonance splits into a peak and a dip. The location of the resonance mode corresponds to the inflection point in the derivative channel. To avoid introducing error in identifying the location of the resonance mode in the $f-B$ map, we do not take the derivative of longitudinal resistance. Most importantly, the location of the resonance is detectable in the longitudinal resistance channel, as shown in Fig.~\ref{fig1} and Fig.~\ref{figHPF}a, providing confirmation for the identified location of the resonance mode.

\subsection{ESR measurement in graphene monolayer and bilayer samples: the absence of resonance response}

\begin{figure*}[t!]
\includegraphics[width=0.75\linewidth]{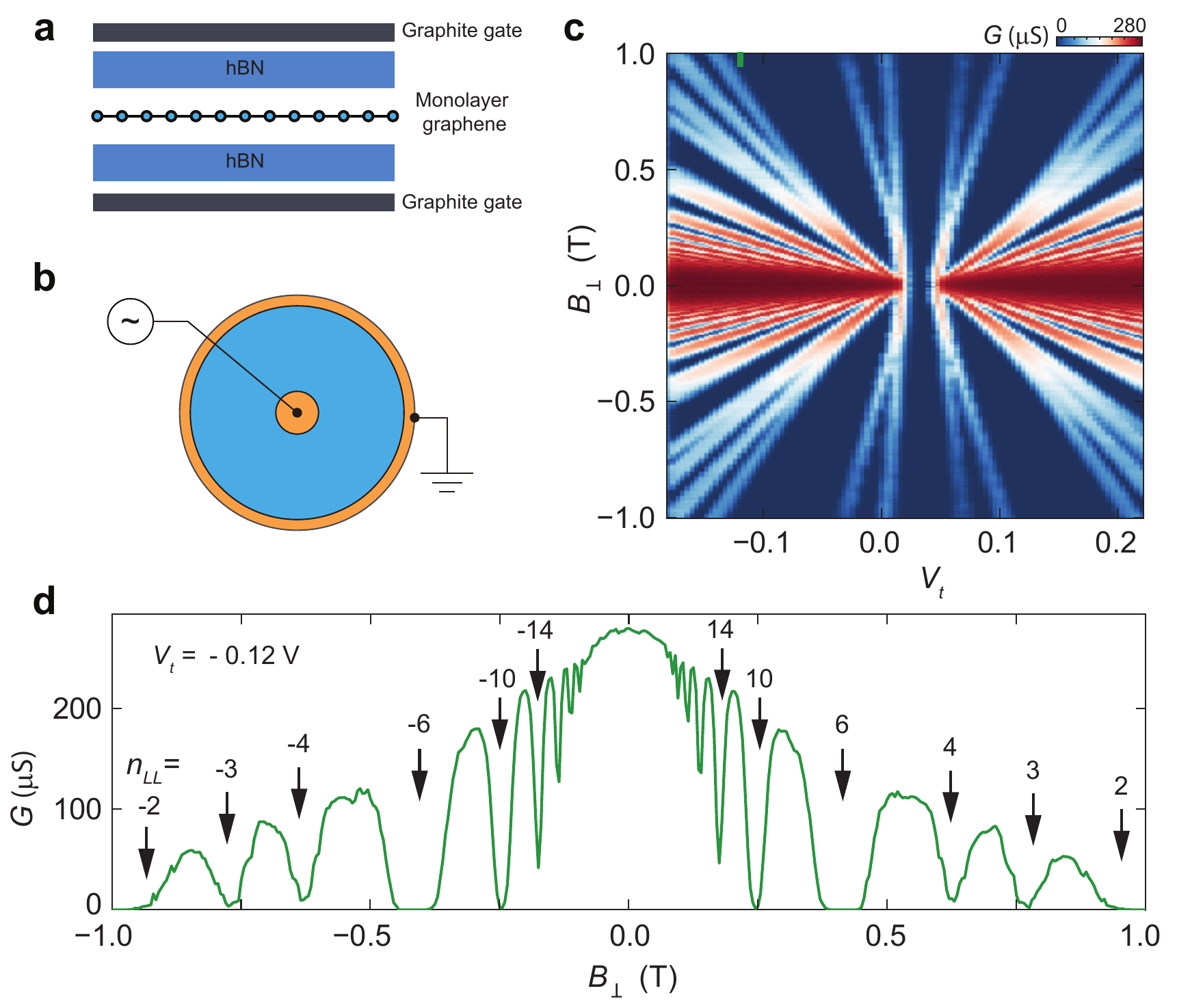}
\caption{\label{fig:corbino} {\bf{Dual-encapsulated monolayer graphene device with Corbino geometry}} 
(a) Schematic of monolayer graphene device dual-encapsulated with hBN and graphite. (b) The edge-free, Corbino geometry where the voltage bias and measured current yield the conductance ~\cite{Zeng2019Corbino,Polshyn2018Corbino}. (c) Magnetotransport measurement of the monolayer Corbino device near the charge neutral point (CNP) at low magnetic field, without any microwave signal. Bulk conductance $G$ is plotted as a function of gate voltage and $B$-field value. (d) Bulk conductance versus out-of-plane magnetic field at $V_t = -0.12$ V, where each discernible quantum Hall effect state or Landau level, $n_{LL}$ is labelled. A large number of quantum oscillation is observed in the magnetic field range of $|B| <1$ T, reflecting the excellent sample quality that is in-line with previous experiments using the same Corbino geometry ~\cite{Zeng2019Corbino,Polshyn2018Corbino}. The Corbino geometry directly probes transport resonse of the sample bulk, without the influence of a more disordered sample edge, giving rise to an enhanced sensitivity in small changes in the charge transport. This is the ideal geometry to look for subtle changes in the transport response induced by potential resonance modes. Notably, the same measurement is performed in samples with Hall-bar geometry with the same result. 
}
\end{figure*}

To accurately identify the resonance response and its underlying mechanism, well-controlled procedures for sample fabrication is needed to minimize potential influence from the substrate and contamination from the fabrication and measurement process. This is particularly crucial given that graphene consists of one layer of atoms.  In order to understand the transport response from graphene, the need to minimize the influence of outside disorder and impurity is well documented in the literature. The advancement in our understanding of the electronic order in graphene and other 2D materials goes hand-in-hand with the progress to isolate graphene from the outside influence ~\cite{Dean.10,Young.12,Lei.13,Zibrov2017,Li.17b,Zeng2019Corbino,Polshyn2018Corbino}. At the same time, a thorough characterization of sample details, such as the width of the electron-hole puddle regime, potential alignment between graphene and the hBN substrate, behavior of quantum oscillation and quantum Hall ferromagnetism provide important clues to support the interpretation of resonance signal. 
Not only are the transport properties of graphene highly susceptible to disorder and impurities from the outside environment, but the resonance response of graphene could be easily modified by the magnetic properties of impurities on the substrate ~\cite{Blick2021}. 
In the following, we show that the microwave resonance response is absent in graphene monolayer and bilayer samples where the influence of substrate impurity and sample disorder is eliminated. These include MLG aligned with hBN substrate, MLG misaligned with hBN substrate, twisted bilayer graphene with large twist angle and AB-stacking Bernal bilayer graphene. There is no flatband physics in these monolayer and bilayer samples. The influence of substrate impurity is minimized using double-encapsulation with graphite and hBN. We also use exfoliated graphene to eliminate the contamination from the process of chemical vapor deposition. These steps ensure ultra-high quality graphene sample, as evidenced by the transport response without microwave radiation. These samples are fabricated with the same procedures compared to the MATBG, allowing us to compare the resonance response between these systems with equal footing. We observe no indication of microwave resonance response.  The absence of microwave resonance in these high quality monolayer and bilayer samples 
is in stark contrast with previously reported resonance behavior in unencapsulated, disordered graphene samples prepared with chemical vapor deposition ~\cite{Blick2020,Blick2021}. As discussed in Ref.~\cite{Sichau2019ESR,Zhang2022thesis}, the resonance response observed in unencapsulated samples may arise from a difference source, such as sample disorder or substrate contamination from the chemical vapor deposition process.

\begin{figure*}[t!]
\includegraphics[width=0.6\linewidth]{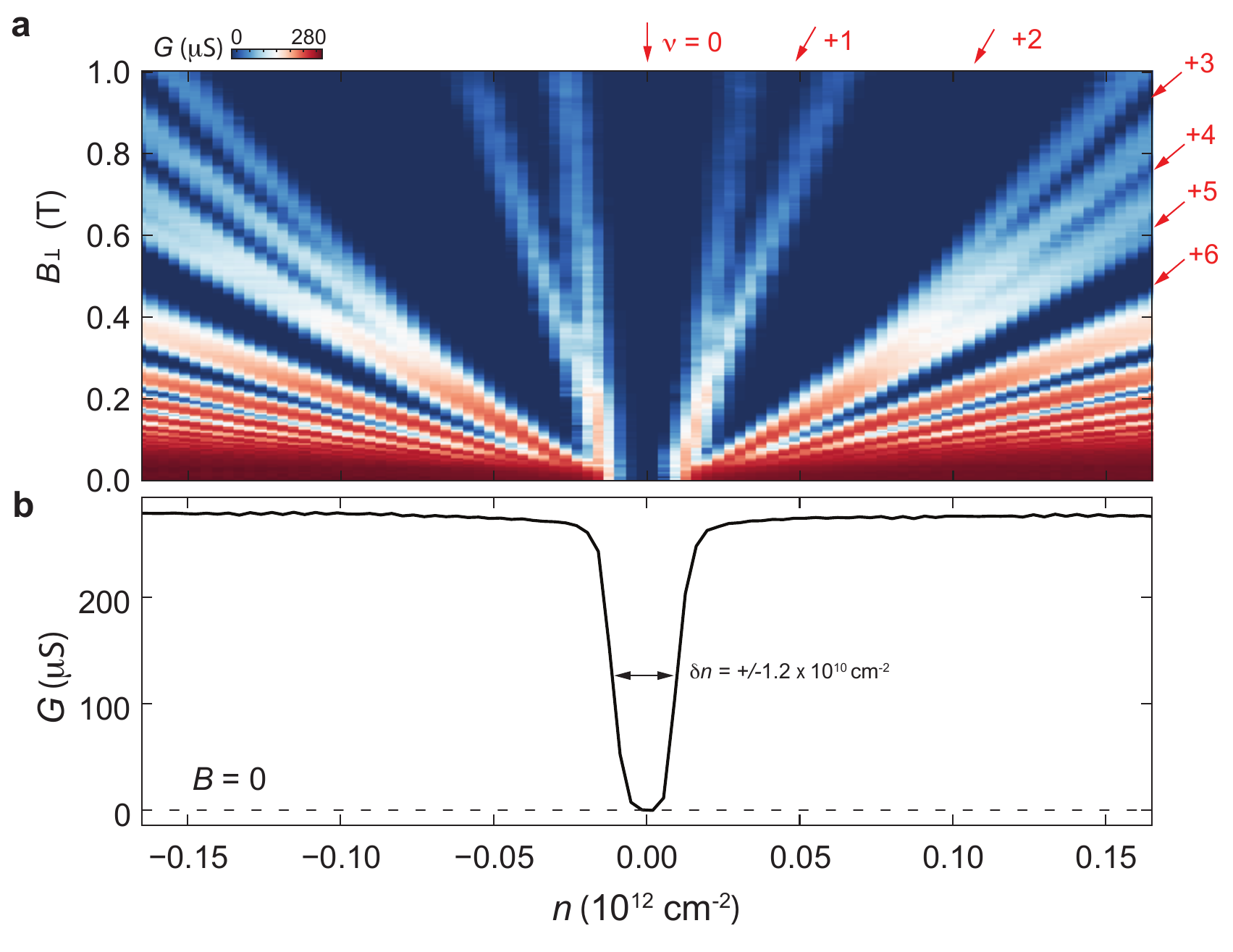}
\caption{\label{fig:CNPMLG} {\bf{Transport measurement of MLG sample, quantum oscillation and gapped Dirac point.}} 
(a) Magnetotransport measurement of the monolayer Corbino device near the charge neutral point (CNP) at low magnetic field, without any microwave signal. Bulk conductance $G$ is plotted as a function of carrier density and $B$-field. Incompressible states are fully developed along integer Landau level filling at low magnetic field of $B < 1$, evidence for excellent sample quality that is in-line with previous experiments ~\cite{Zeng2019Corbino,Polshyn2018Corbino}. (b) Bulk condutance as a function of carrier density measured at $B = 0$. Both panels show that the sample is fully insulating at the CNP at $B=0$, and this is strong indication of alignment between graphene and the hBN substrate ~\cite{Polshyn2018Corbino}. The density range of the charge fluctuation regime is $\delta n = \pm 1.2 \times 10^{10}$ cm$^{-2}$, which is comparable with the MATBG sample studied in the main text.
}
\end{figure*}

Fig.~\ref{fig:corbino}a-b  shows the schematic of a monolayer graphene sample with dual hBN and graphite encapsulation. Double-encapsulation ensures high sample quality, so that the impact of charge fluctuation and outside impurities on transport response is minimized ~\cite{Li.17b,Zibrov2017}. The MLG sample is further shaped into the Corbino geometry, which reduces the influence of disorder on the sample edge ~\cite{Zeng2019Corbino,Polshyn2018Corbino}. The combination of double-encapsulation and the Corbino geometry provides excellent sensitivity in transport measurement to subtle changes in the sample resistivity, which is ideal for detecting potential resonance response.

The excellent sample quality is demonstrated by magneto-transport measurement in Fig.~\ref{fig:corbino}c-d. The Landau fan around the CNP exhibits fully developed incompressible states at every integer filling, from $\nu_{LL}=1$ to $\nu_{LL}=6$, at a relatively low magnetic field of $B = 1$ T (Fig.~\ref{fig:corbino}c). At the same time, Fig.~\ref{fig:corbino}d shows an abundance of quantum oscillation when varying the $B$-field at a fixed carrier density. Notably, the sample is almost fully insulating in the absence of an external magnetic field (Fig.~\ref{fig:CNPMLG}). This is strong evidence for a robust energy gap at the Dirac point of MLG due to the alignment between MLG and the hBN substrate ~\cite{Polshyn2018Corbino,Zibrov_evendemoninator}. 
A series of experiments have shown that such alignment removes the valley degrees of freedom, which is equivalent to the sublattice in the zeroth Landau level, at low magnetic field ~\cite{Hunt2013hofstadter,Zibrov_evendemoninator,Polshyn2018Corbino}. According to the comprehensive analysis in Ref.~\cite{Zibrov_evendemoninator} examining the energy gap of the CNP and even-denominator fractional quantum Hall effect state, a sublattice splitting $\Delta_{AB}$ stabilizes a valley-polarized charge density wave at low magnetic field as the robust ground state near the CNP of MLG, where all electrons occupy a single valley.

Fig.~\ref{fig:MWfans}e-h plots the high-pass filter of sample conductance with magnetic field, in the density-magnetic field ($n-B$) maps measured at different microwave frequency. This is the same measurement as shown in Fig.~\ref{figN}c. We have saturated the color scale so that small variations in the sample conductance will be visible. In each panel, we mark the expected location of $r_1$ and $r_2$ for the corresponding microwave frequency. A microwave-induced resonance is expected to emerge as a horizontal feature similar to the observation in  Fig.~\ref{figN}c. All transport features can be accounted for by quantum Hall effect states emanating from the CNP and no resonance response can be identified.

\begin{figure*}
\includegraphics[width=1\linewidth]{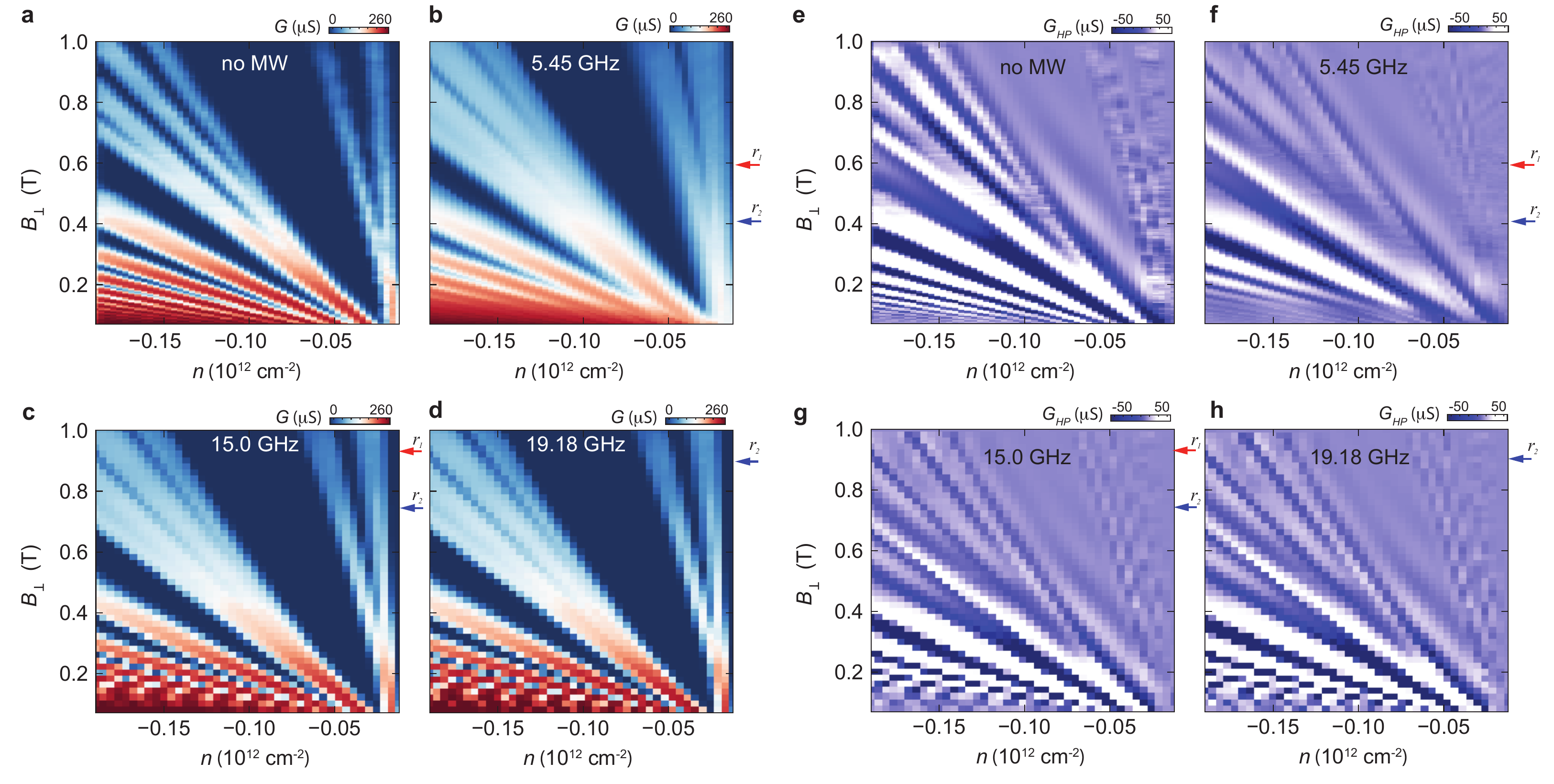}
\caption{\label{fig:MWfans} (a-d) Bulk conductance $G$ as a function of carrier density $n$ and magnetic field $B$ measured at different microwave frequency.  Large microwave power is used here to enhance the signal of the potential resonance response.  (e-h) High-pass filter is applied to eliminate the slow changing background in the bulk conductance as a function of $B$-field, which is comparable to $dG/dB$. After high-pass filter, $G_{HP}$ is plotted as a function of carrier density $n$ and magnetic field $B$ measured at different microwave frequency. 
Horizontal arrows mark the expected position of $r_1$ and $r_2$ modes for each microwave frequency. Potential resonance mode is expected to induce changes in the sample conductance at the $B$-field values near the arrows. We do not observe any indications of resonance response. The only influence of microwave radiation is to reduce the magnitude of transport response at each integer Landau level filling, which results from an elevated electron temperature from microwave-induced heating. 
}
\end{figure*}

Fig.~\ref{fig:monoline} plots the sample conductance as a function of magnetic field measured at a fixed carrier density with different microwave frequencies. While sample conductance in panel (b) exhibits a series of oscillations, the locations of these features are insensitive to varying microwave frequency. This, again, points towards the absence of microwave induced resonance.

\begin{figure*}
\includegraphics[width=0.7\linewidth]{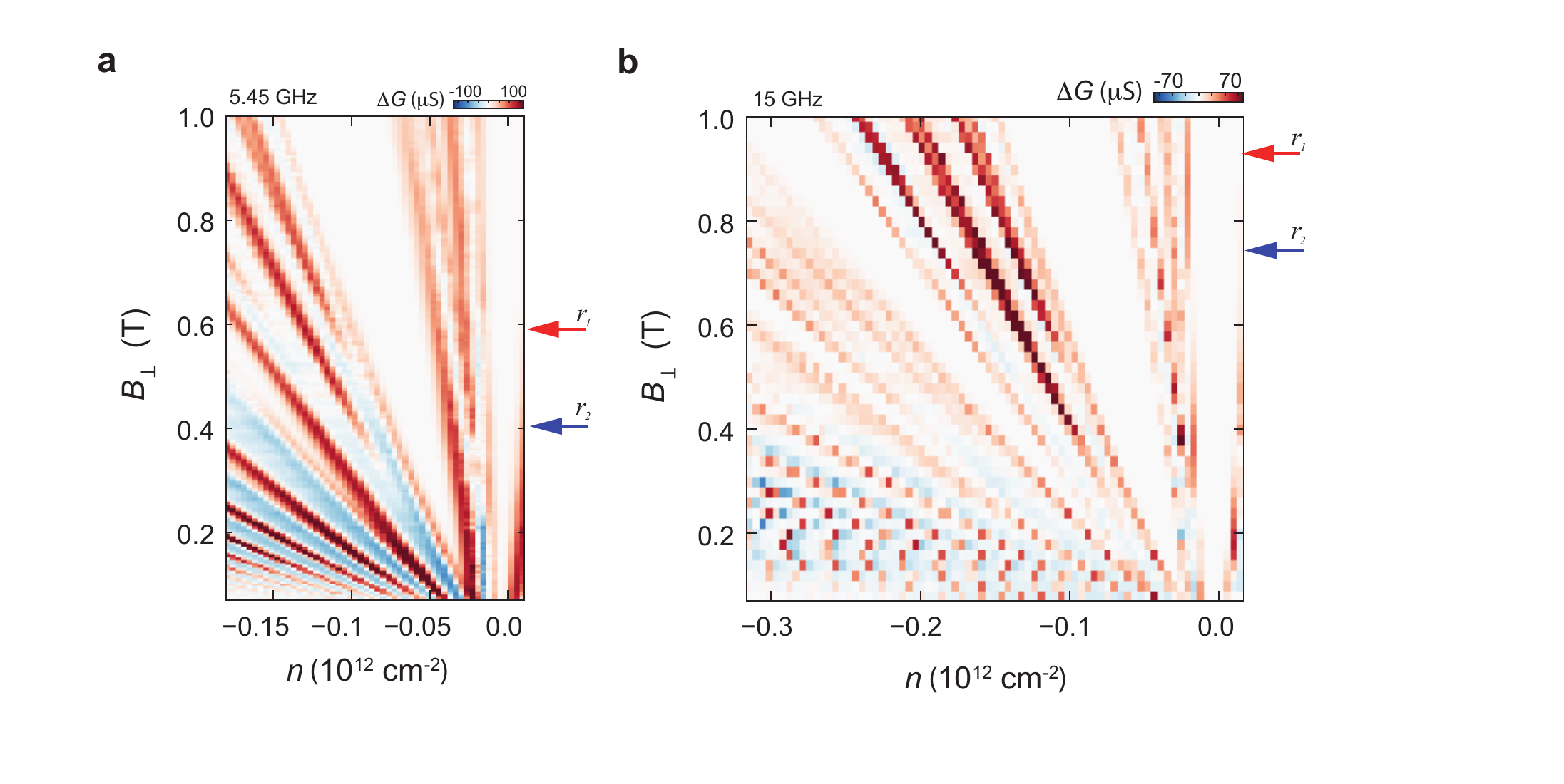}
\caption{\label{fig:MWdif}  Changes in bulk conductance, $\Delta G$, induced by microwave radiation at (a) $f = 5.45$ GHz and (b) $15$ GHz. The horizontal red and blue arrows mark the expected location of the $r_1$ and $r_2$ modes for the microwave frequency. Prominent response in $\Delta G$ is observed near the edge of the Landau levels. Since transport response near the edge of the Landau level is highly sensitive to changes in electron temperature, this observation is consistent with microwave-induced heating in the electron temperature. Most importantly, a ferromagnetic or paramagnetic resonance mode is expected to emerge as prominent changes in $\Delta G$ near the $B$-field value marked by the blue arrow, whereas a resonance mode with $J_H > 0$ will appear near the red arrow as a signal in $\Delta G$. There is no detectable transport response that matches the expectation of a microwave-induced resonance.  } 
\end{figure*}

\begin{figure*}
\includegraphics[width=0.6\linewidth]{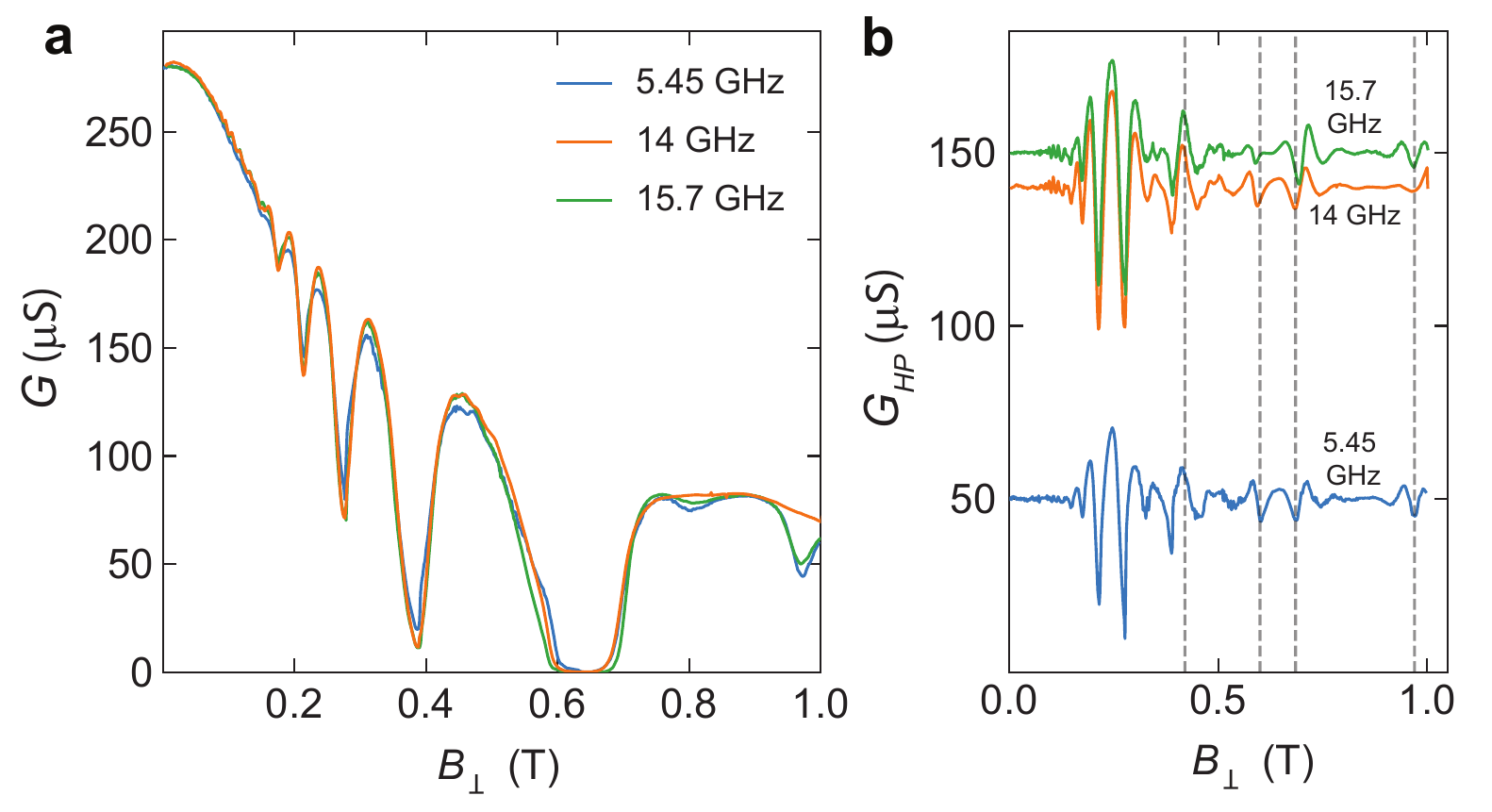}
\caption{\label{fig:monoline} 
(a) Bulk conductance as a function of magnetic field $B$ measured with different microwave frequency. Transport response shows no frequency dependence. (b) The same bulk conductance as in panel (a), after high-pass filter, allowing us to identify weaker features in the bulk conductance. No dependence on microwave frequency is observed. Taken together, we do not observe microwave-induced resonance in MLG.
}
\end{figure*}

Fig.~\ref{fig:MWdif} plots the difference in sample conductance with and without microwave radiation, $\Delta G$. While microwave induces substantial changes in sample conductance near the edge of the Landau level, two observations testify that these changes are due to an increase in electron temperature resulting from the high power of microwave radiation, instead of a resonance response. First, $\Delta G$ is not confined near any specific magnetic field values, despite the well-defined microwave frequency. This is inconsistent with the expected behavior of spin resonance. Second, sample conductance near the edge of the Landau level is the most sensitive to changes in electron temperature. The map of $\Delta G$ can be perfectly simulated by taking the difference between measurements performed at different temperature. Taken together, our measurements on MLG indicate that microwave radiation on high quality graphene samples only induces heating that enhances the electron temperature. There is no detectable changes in sample transport that is indicative of a potential resonance response. 

We have also performed microwave measurement on a MLG sample misaligned with the hBN substrate (Fig.~\ref{fig:MLG2}). There is no indication of microwave-induced resonance response.
The lack of resonance response in MLG samples are in excellent agreement with our findings in the IU regime of MATBG, where no resonance response is observed. 

Microwave resonance measurements performed in tBLG sample with large twist angle, as well as Bernal bilayer sample with AB stacking is shown in Fig.~\ref{figMWMSU} and Fig.~\ref{figBernal}.

\begin{figure*}
\includegraphics[width=0.85\linewidth]{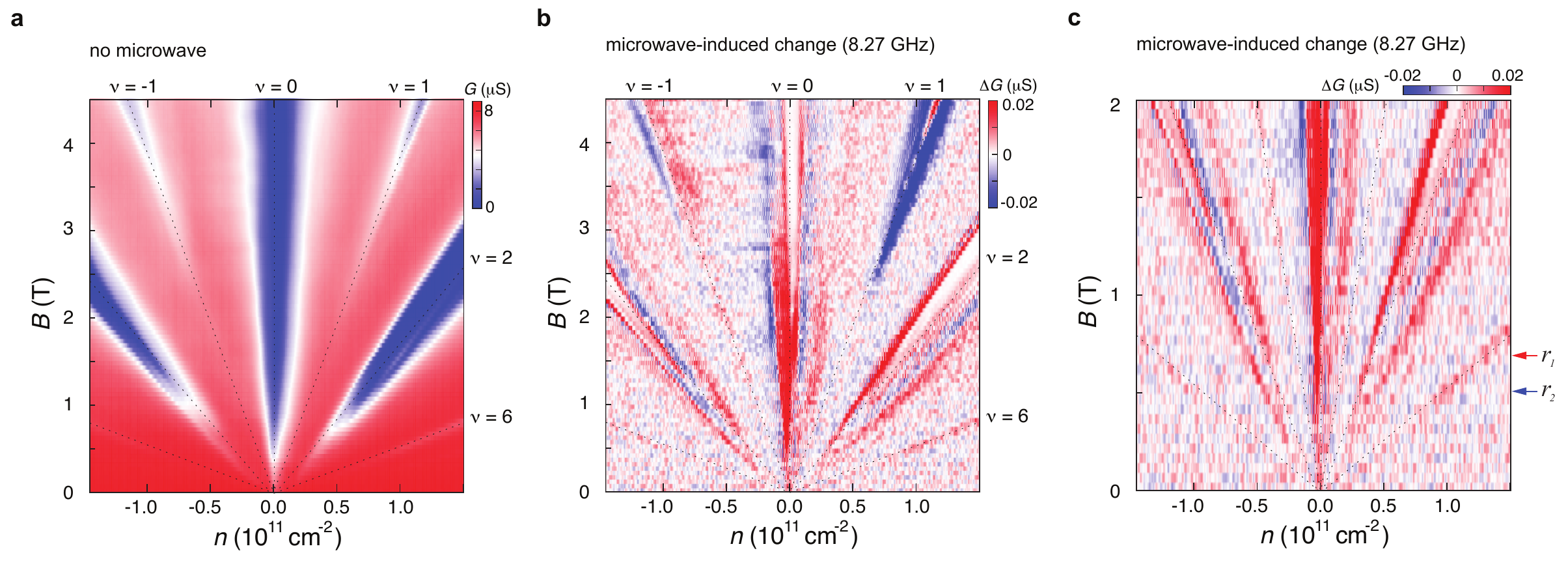}
\caption{\label{fig:MLG2} {\bf{Microwave resonance measurement in a second MLG sample.}} The same microwave resonance is performed in a second MLG sample where MLG is misaligned with the hBN substrate. Such misalignment is evidenced by the conductive behavior of the Dirac peak at low $B$. 
(a) Bulk conductance as a function of carrier density $n$ and magnetic field $B$ measured without microwave radiation. (b) Radiation of microwave at frequency of $f=8.27$ GHz induces changes in the bulk conductance measurement $\Delta G$. Panel (b) plots $\Delta G$ as a function of carrier density $n$ and magnetic field $B$. The color scale is saturated to reveal the slightest changes in bulk conductance. Microwave-induced changes appear near the edge of the Landau levels, where transport measurement has strong temperature dependence. At $f = 8.27$ GHz, a resonance mode with $g=2$ is expected to appear at $B < 1.5$ T. The fact that microwave-induced response persist to high magnetic field up to $B = 4.5$ T points towards an origin in microwave-induced heating. (c) Zoom in of panel (b) in the low-field regime. Red and blue arrows mark the expected position of $r_1$ and $r_2$ modes. There is no indication of resonance behavior at, or near, these positions. 
This MLG sample shows far less quantum oscillation compared to the sample in Fig.~\ref{fig:corbino}. The absence of resonance response, therefore, indicates that the lack of resonance response is not due to the presence of quantum oscillation. This is also consistent with our observation in MATBG. An abundance of quantum oscillation is observed in the IF$_2$ regime without microwave radiation. These quantum oscillations are replaced by the resonance response in the presence of microwave radiation (see Fig.~\ref{fig:LL}). 
}
\end{figure*}

\clearpage

\begin{figure*}[t]
\includegraphics[width=0.75\linewidth]{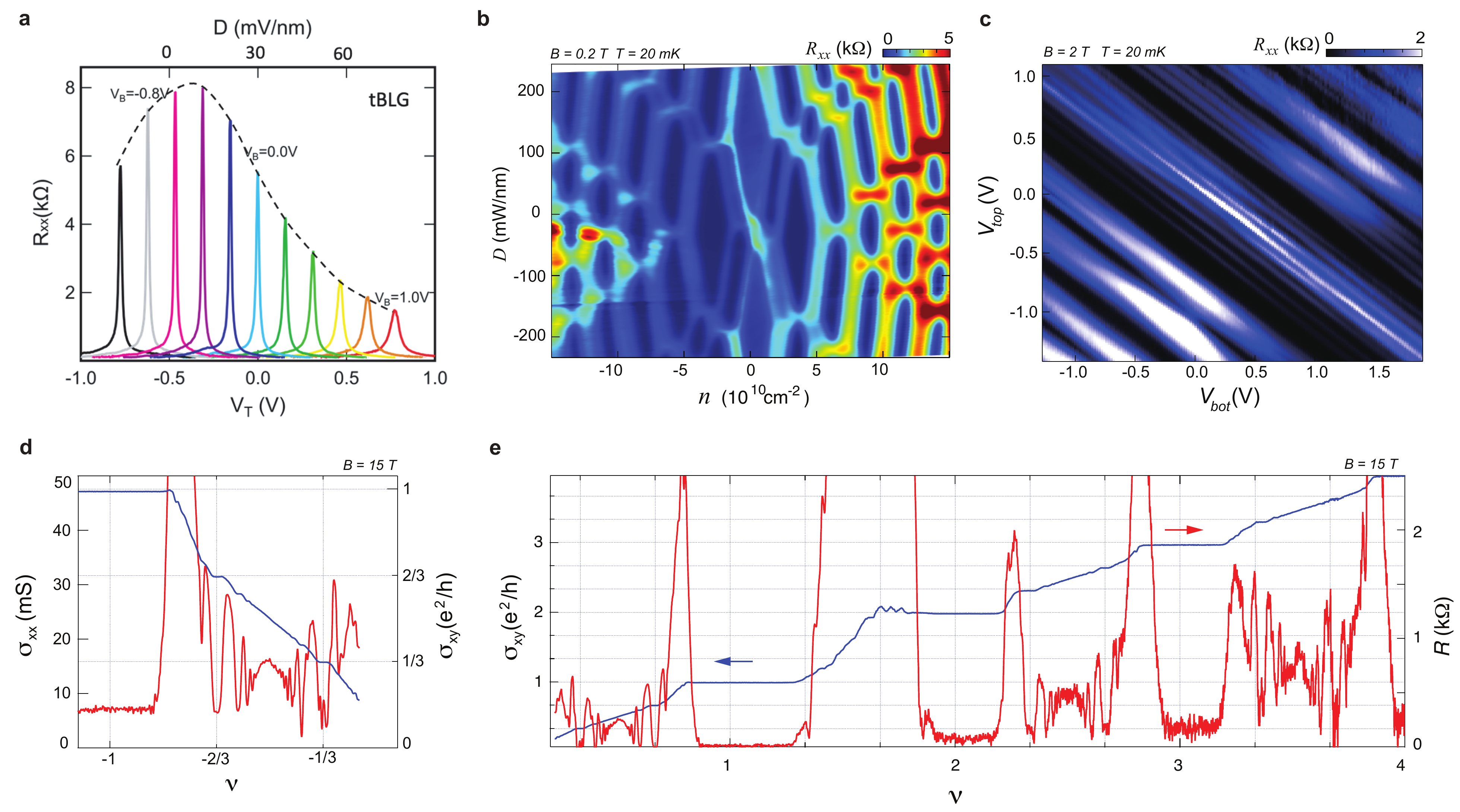}
\caption{\label{figtBLG} {\bf{Dual-encapsulated twisted bilayer graphene device with large twist angle}} 
The following transport measurements show that two graphene layers are rotationally misaligned with a large twist angle. (a) The resistance peak at the CNP decreases with increasing displacement field. (b) In a weak magnetic field of $B = 0.2$ T, top and bottom graphene layers are decoupled. Varying the perpendicular electric field induces a series of transitions, where occupied Landau levels shift from one layer to another. This is similar to the behavior shown in Ref.~\cite{Sanchez2017}. (c) Gate-gate map of longitudinal resistance measured at $B = 2$ T. The $N=0$ Landau level splits into $8$ diagonal lines, which correspond to $8$ symmetry breaking Landau levels. The $8$-fold degeneracy arises from spin, valley and layer degrees of freedom. (d-e) Fractional quantum Hall effect measured at $B = 15$ T. The sequence of fractional quantum Hall effect shows excellent agreement with the $N=0$ Landau wavefunction across a wide density range, $-1< \nu_{LL} < +4$. This confirms the 8-fold degeneracy that includes layer as a quantum number. At the same time, the FQHE sequence in the density range of $-1 <\nu_{LL} < 0$ is distinctly different from that of Bernal bilayer graphene ~\cite{Zibrov2017,Li.17b}. Most importantly, the transport behavior of this tBLG shows no sign of flatband physics, confirming the large twist angle between top and bottom graphene layers.
}
\end{figure*}

\begin{figure*}[b!]
\includegraphics[width=0.9\linewidth]{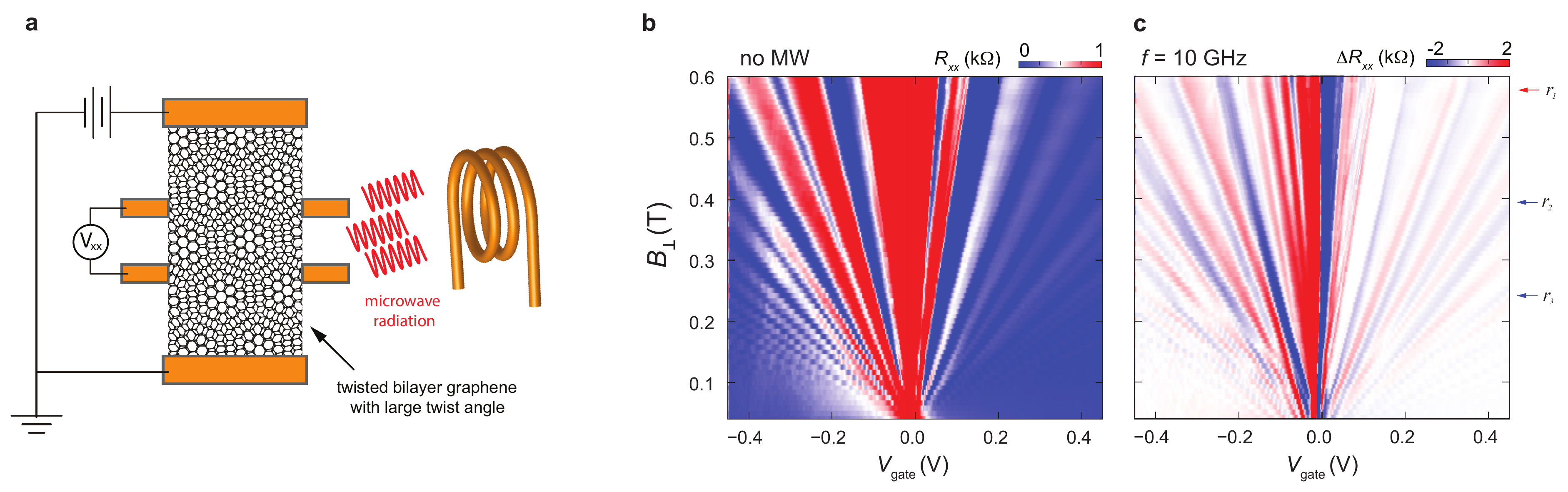}
\caption{\label{figMWMSU} {\bf{Resonance measurement in tBLG sample with large twist angle}} (a) Schematic of resistively-detected microwave resonance measurement. Microwave photon is radiated onto the sample using a pick up coil. More details of the microwave setup can be found in the Appendix A of Ref.~\cite{Zhang2022thesis}. (b) Longitudinal resistance as a function of magnetic field and gate voltage measured without microwave radiation. The main sequence of integer quantum Hall effect states corresponds to Landau level filling of $\nu_{LL}= \pm4$, $\pm8$ .... which is in excellent agreement with the expected behavior of twisted bilayer graphene with large twist angle. (c) The $V-B$ map of $\Delta R_{xx}$, which corresponds to microwave-radiation-induced changes in \Rxx. Microwave frequency is set at $f = 10$ GHz. Horizontal arrows mark the expected position for resonance mode $r_1$, $r_2$ and $r_3$. A potential resonance response is expected to be localized near certain magnetic field values, as the photon energy matches an energy gap that depends on the Zeeman coupling.  Instead, we find that microwave-induced changes in \Rxx\ track constant Landau level filling and remains detectable over the entire magnetic field range. Such response arises from a microwave-induced increase in the electron temperature.  Notably, the result in panel (c) is consistent with Fig.~\ref{fig:MWdif} and Fig.~\ref{fig:MLG2}. Taken together, there is no indication of resonance responses in graphene monolayer and bilayer samples without flatband physics. 
}
\end{figure*}

\begin{figure*}[h]
\includegraphics[width=0.6\linewidth]{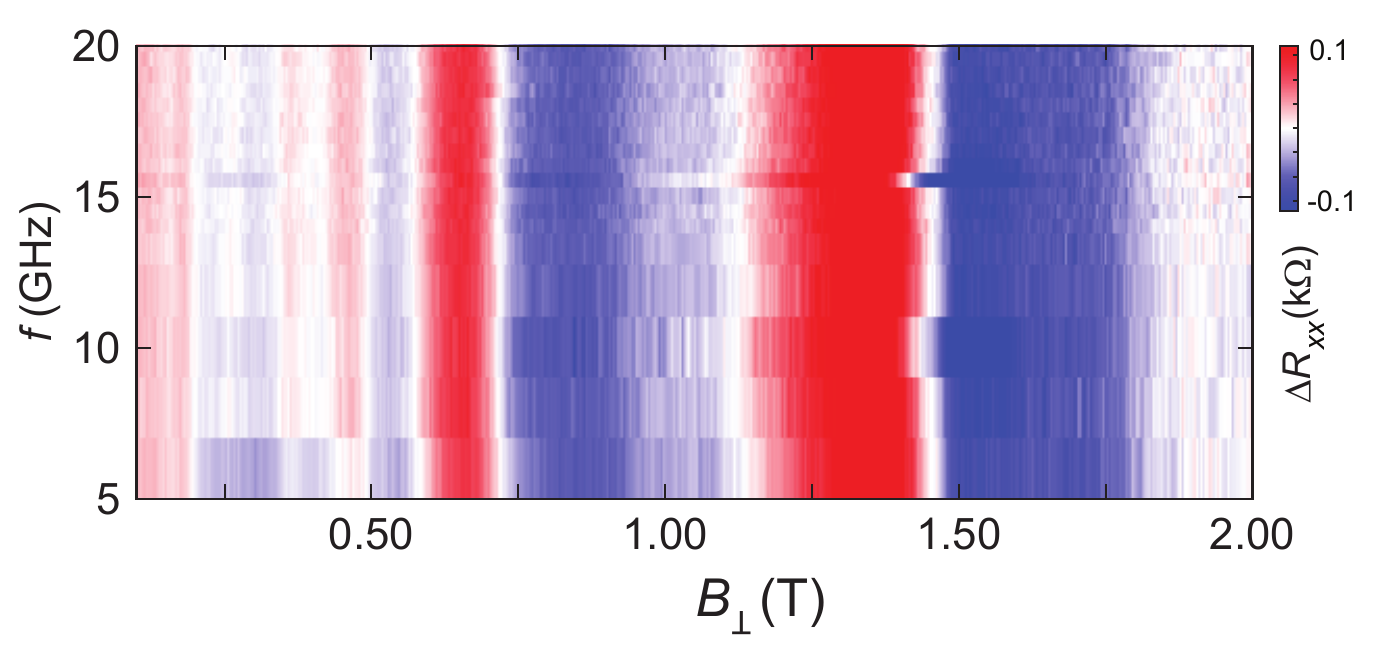}
\caption{\label{figBernal} {\bf{Resonance measurement in a Bernal BLG sample with AB-stacking order.}} Microwave-induced change in the longitudinal resistance, $\Delta R_{xx} = R_{xx}$(on)$-R_{xx}$(off), as a function of magnetic field \Bperp\ and microwave frequency. Here $R_{xx}$(on) ($R_{xx}$(off)) denotes longitudinal resistance measured with (without) microwave radiation. No frequency dependence is observed in transport response, suggesting that there is no observable resonance response in the sample.
}
\end{figure*}

\begin{table}[h]
\begin{center}
\caption{We summarize the result of microwave resonance measurement in different samples.
}
\label{DiodeEffect2}
\begin{ruledtabular}
 \begin{tabular}{ccc} 
 \\
\multicolumn{1}{c}{} & \multicolumn{1}{c}{NOTE} & \multicolumn{1}{c}{resonance response}   \\  
\\
\hline
\\
MATBG & Near the magic angle & Prominent resonance response is observed in the IF$_2$ regime. \\

 & & No resonance in the IU regime     \\
\\
 \hline 
\\
MLG sample 1 & graphene/hBN alignment & no resonance   \\
\\
 \hline
\\
MLG sample 2 & graphene/hBN misaligned & no resonance     \\
\\
  \hline
\\
Twisted bilayer graphene & large twist angle & no resonance    \\
 &  no flatband or Dirac revival&
\\
\\
    \hline
    \\
Bernal bilayer graphene & doubly encapsulated & no resonance    \\  
\\
 \end{tabular}
\end{ruledtabular}
\end{center}
\end{table}

\clearpage

\section{Supplementary Text: Magnon spectra and microwave resonance}
Motivated by the observation that the resonance features are present and mostly unaltered in a significant range of doping in Fig.~\ref{figN}, which involves both the insulating resistive states around integer filling as well as metallic regions, we assume that the resonance features are associated with the high-temperature reset orders \cite{Zondiner2020cascade,Wong2020cascade} rather than the correlated insulators. With the exception of $\nu=1$, prominent resonance features are primarily present in the range $2\leq |\nu| \leq 3$ such that we next focus on the reset orders that set in at $|\nu|=2$.
In Ref.~\cite{Christos2020C2T}, possible high-temperature reset order parameters at $|\nu|=2$ were classified. Out of all 15 candidate orders, only the following six,
\begin{equation}
    \vec{s}, \quad \tau_z\vec{s}; \quad (\mu_x,\mu_y)\vec{s}, \quad \tau_z(\mu_x,\mu_y)\vec{s}; \quad \rho_y(\tau_x,\tau_y)\vec{s}; \quad \rho_y(\mu_x,\mu_y)(\tau_x,\tau_y)\vec{s}, \label{CandidateOrders} 
\end{equation}
transform non-trivially under SU(2)$_s$ rotation as required by the $g=2$ slopes of our microwave resonance frequencies while not giving rise to additional Fermi surfaces \cite{Christos2020C2T,Zondiner2020cascade}. In \equref{CandidateOrders}, $s_j$, $\tau_j$, $\mu_j$, and $\rho_j$ are Pauli matrices in spin, valley, mini-valley, and sublattice space. The first ($\vec{s}$, $\tau_z\vec{s}$) and second ($\mu_{x,y}\vec{s}$, $\tau_z\mu_{x,y}\vec{s}$) pair of states in \equref{CandidateOrders} are each ``Hund's partners'' as they are degenerate in the limit where the intervalley Hund's coupling $J_H$ is neglected, due to the resulting SU(2)$_+\times$ SU(2)$_-$ spin symmetry. The Hund's partner of the three remaining states are omitted in \equref{CandidateOrders} as they do not carry spin and therefore do not couple to the Zeeman field. We will here take $J_H$ to be finite in the analysis but allow for it to be potentially small. However, the energy scales quantifying how `close' the system's symmetries are to larger continuous symmetry groups, are assumed to be large for our purposes. This follows from the fact that the associated pseudo-Goldstones modes have gaps of order of a few meV \cite{Khalaf2020collective} and thus far beyond the frequency range studied in our experiments. Together with the valley-charge-conservation symmetry, U(1)$_v$, we assume a continuous symmetry group of U(1)$_v\times$SU(2)$_+\times$ SU(2)$_-$, which is (weakly) broken down to U(1)$_v\times$SU(2)$_s$ by $J_H$.

\subsection{Dirac revival from spin polarization and resonance frequencies}

Before addressing the remaining options, let us begin by analyzing the first two states in \equref{CandidateOrders}, which correspond to parallel spin polarization in the two valleys and its ``Hund's partner'', a state with anti-parallel spins in opposite valleys. Being degenerate in the limit $J_H\rightarrow 0$, we have to study both states simultaneously. To this end, we consider a phenomenological triangular-lattice spin model, involving the spin-$1/2$ operators $\hat{\vec{S}}_{j}^\tau$ with site index $j$ and valley index $\tau=\pm$. The Hamiltonian reads as
\begin{equation}
    \hat{H}_{H} = -\sum_j\sum_{\tau=\pm} \sum_{\mu} J_\mu \hat{\vec{S}}_{j,\tau} \cdot \hat{\vec{S}}_{j+\vec{\eta}_\mu,\tau} + J_H \sum_j \hat{\vec{S}}_{j,+} \cdot \hat{\vec{S}}_{j,-} -  g \mu_B \sum_j \sum_{\tau=\pm} \vec{B}\cdot \hat{\vec{S}}_{j,\tau}, \label{ToyModel}
\end{equation}
where $J_\mu$ are the intra-valley exchange coupling parameters associated with the triangular-lattice bond vector $\vec{\eta}_\mu$, which we assume to respect $C_{3}$ symmetry, and $J_{H}$ is the inter-valley Hund's interaction, which determines whether the system favors parallel (ferromagnetic, $J_H<0$, expected for Coulomb interactions \cite{Chatterjee2020}) or anti-parallel (anti-ferromagnetic, $J_H>0$) spins in the two valleys. The last term in \equref{ToyModel} is the coupling to the Zeeman field $\vec{B}$, with Land\'e factor $g=2$ and Bohr magneton $\mu_B$. For clarity, we indicate operators, like $\hat{\vec{S}}_{j,\tau}$, with a `hat'.

To compute the magnon spectrum of this model, we follow previous works \cite{PhysRevLett.91.017205} and express the spin operators in terms of its component, $\hat{S}_{j,\parallel}^\tau$, along and its two components, $\hat{\vec{S}}_{j,\perp}^\tau$, perpendicular to the direction $\vec{m}_{j,\tau}$ of the spin configuration stabilized by \equref{ToyModel} in the classical limit (formally defined as the minimum of $E_{\text{cl}}[\vec{m}_{j,\tau}]=\hat{H}_{H}|_{\hat{\vec{S}}_{j}^\tau\rightarrow S \vec{m}_{j,\tau}}$, with $\vec{m}^2_{j,\tau}=1$). As we are interested in magnetic states that preserve moir\'e translational symmetry, we here focus on $\vec{m}_{j,\tau} = \vec{m}_{\tau}$, which will be realized if, e.g., $J_\mu>0$. The decomposition then takes the form
\begin{equation}
    \hat{\vec{S}}_{j,\tau} =  \vec{m}_{\tau} \hat{S}_{j,\tau}^\parallel  + \hat{\vec{S}}_{j,\tau}^\perp, \quad \vec{m}_{\tau}\cdot \hat{\vec{S}}_{j,\tau}^\perp =0,\quad \vec{m}_{\tau}^2=1.
\end{equation}
Inserting this into \equref{ToyModel}, we get 
\begin{align}\begin{split}
    \hat{H}_{H} = &-\sum_j\sum_{\tau=\pm} \sum_{\mu} J_\mu \hat{S}_{j,\tau}^\parallel \cdot \hat{S}_{j+\vec{\eta}_\mu,\tau}^\parallel + J_H (\vec{m}_+\cdot \vec{m}_-) \sum_j \hat{S}_{j,+}^\parallel \hat{S}_{j,-}^\parallel -  g \mu_B \sum_j \sum_{\tau=\pm} (\vec{B}\cdot \vec{m}_\tau) \hat{S}_{j,\tau}^\parallel \\ 
    &-\sum_j\sum_{\tau=\pm} \sum_{\mu} J_\mu \hat{\vec{S}}_{j,\tau}^\perp \cdot \hat{\vec{S}}_{j+\vec{\eta}_\mu,\tau}^\perp + J_H \sum_j \hat{\vec{S}}_{j,+}^\perp \cdot \hat{\vec{S}}_{j,-}^\perp + \Delta H
\label{HamiltonianSplitIntoTwoParts}\end{split}\end{align}
where 
\begin{equation}
    \Delta H = \sum_j\sum_{\tau=\pm} \hat{\vec{S}}^\perp_{j,\tau} \cdot \left[ J_H \vec{m}_{-\tau} S_{j,-\tau}^\parallel - \mu_B g\vec{B}\right]. \label{FormOfDeltaH}
\end{equation}
We next generalize to spin-$S$ operators and employ a Holstein-Primakoff transformation \cite{HPTransformation}. To that end, let us choose the unit vectors $\vec{e}_{a,\tau}$ such that $\{\vec{e}_{1,\tau},\vec{e}_{2,\tau},\vec{m}_\tau\}$ form an orthonormal, right-handed (i.e., $\vec{m}_\tau=\vec{e}_{1,\tau} \times \vec{e}_{2,\tau}$) triad and write
\begin{equation}
    \hat{S}_{j,\tau}^\parallel = S - \hat{b}_{j,\tau}^\dagger \hat{b}_{j,\tau}^\pdagger, \quad \hat{\vec{S}}^\perp_{j,\tau} = \sum_{s=\pm} \vec{e}_\tau^s \hat{S}_{j,\tau}^s, \quad \vec{e}_\tau^\pm = \frac{1}{2}\left( \vec{e}_{1,\tau} \mp i \vec{e}_{2,\tau}\right), \quad \hat{S}_{j,\tau}^+ = (\hat{S}_{j,\tau}^-)^\dagger = \sqrt{2S - \hat{b}_{j,\tau}^\dagger \hat{b}_{j,\tau}^\pdagger} \hat{b}_{j,\tau}^\pdagger,
\end{equation}
where $\hat{b}_{j,\tau}$ ($\hat{b}_{j,\tau}^\dagger$) are bosonic annihilation (creation) operators. Plugging this into \equref{HamiltonianSplitIntoTwoParts} and expanding up to subleading order in $1/S$ (treating $\vec{B}$ of order $S$), we get
\begin{align}\begin{split}
    \hat{H}_{H} &=  E_{\text{cl}}[\vec{m}_\tau] - S \sum_{j,\tau,\mu} J_\mu \left( \hat{b}_{j,\tau}^\dagger \hat{b}_{j+\vec{\eta}_\mu,\tau}^\pdagger + \text{H.c.} \right) - \sum_{j,\tau} \mu_\tau \hat{b}_{j,\tau}^\dagger \hat{b}_{j,\tau}^\pdagger \\ & \quad +  \sum_j \left(t \, \hat{b}_{j,+}^\dagger \hat{b}_{j,-}^\pdagger + \Delta\, \hat{b}_{j,+}^\dagger \hat{b}_{j,-}^\dagger + \text{H.c.}\right) +\mathcal{O}(1/S^0) + \Delta H,
\end{split}\end{align}
where 
\begin{equation}
    \mu_\tau = - 2 S\sum_\mu J_\mu + J_H S (\vec{m}_+\cdot \vec{m}_-) - g\mu_B\vec{B}\cdot \vec{m}_\tau , \quad t = 2S J_H \vec{e}_+^-\cdot \vec{e}_-^+, \quad \Delta = 2S J_H \vec{e}_+^- \cdot \vec{e}_-^-
\end{equation}
and
\begin{equation}
    E_{\text{cl}}[\vec{m}_\tau] = N S^2 \left[ - 2 \sum_{\mu} J_\mu + J_H \, \vec{m}_+\cdot \vec{m}_- - S^{-1}g \mu_B \sum_\tau \vec{B}\cdot \vec{m}_\tau \right] \label{ClassicalEnergy}
\end{equation}
is just the classical energy of magnetic order with $\vec{m}_{j,\tau} = \vec{m}_{\tau}$.

In the limit $S\rightarrow \infty$, $\vec{m}_\tau$ is determined by minimization of \equref{ClassicalEnergy}. At the minimum, $g\mu_B\vec{B} - S J_H \vec{m}_{-\tau} \propto \vec{m}_\tau$. Consequently, $\Delta H$ in \equref{FormOfDeltaH} is $\mathcal{O}(1/S^0)$ and can be neglected in our large-$S$ analysis.

Let us first consider a ferromagnetic Hund's coupling, $J_H < 0$. In that case, we get $\vec{m}_+ = \vec{m}_- = \vec{B}/|\vec{B}|$ from \equref{ClassicalEnergy}; we can thus choose $\vec{e}_{i,+}=\vec{e}_{i,-}$ and get $\Delta=0$ and $t=SJ_H$. It is straightforward to diagonalize the resulting form of $\hat{H}_{H}$ by Fourier transformation and introducing new bosonic operators $\hat{\gamma}_{\vec{k},\pm}$,
\begin{align}
    \hat{H}_{H} &= \text{const.} + \sum_{\vec{k},\tau} \left[ (\xi_{\vec{k}} + g \mu_B |\vec{B}| - SJ_H)  \hat{b}^\dagger_{\vec{k},\tau}\hat{b}^\pdagger_{\vec{k},\tau} + S J_H \hat{b}^\dagger_{\vec{k},\tau}\hat{b}^\pdagger_{\vec{k},-\tau} \right] + \mathcal{O}(1/S^0), \quad \xi_{\vec{k}} = 2S\sum_\mu J_\mu (1-\cos \vec{\eta}_\mu\cdot\vec{k}) \\
    &= \text{const.} + \sum_{\vec{k},p=\pm} E_{\vec{k},p} \hat{\gamma}^\dagger_{\vec{k},p}\hat{\gamma}^\pdagger_{\vec{k},p} + \mathcal{O}(1/S^0),  \quad E_{\vec{k},p} = \xi_{\vec{k}} + S |J_H| (1-p) + g \mu_B |\vec{B}|.  \label{FerromagneticResult}
\end{align}
Due to the two valleys, there are two distinct magnon modes: one of them ($p=+$) is the Goldstone mode associated with the spontaneously broken spin-rotation symmetry. Therefore, its gap $g\mu_B |\vec{B}|$ vanishes in the limit $\vec{B}\rightarrow 0$. At small $\vec{k}$, the spectrum is quadratic, $E_{\vec{k},+} \sim J \vec{k}^2 a^2$, where $a$ is the moir\'e lattice constant and $J = \frac{S}{2}\sum_\mu (\vec{\eta}_\mu/a)^2$, a measure of the spin stiffness. This is consistent with general expectations \cite{CountingGoldstones}: among the three generators, $\hat{\Lambda}_{i} = \sum_{j,\tau} (\hat{\vec{S}}_{j,\tau})_i$, $i=x,y,z$, one (linear combination) has a finite expectation value in the ground state and hence $\rho_{i,i'} \propto \langle{[\hat{\Lambda}_{i},\hat{\Lambda}_{i'}]}\rangle$ has rank two. With the two broken generators ($n_{\text{BG}}=2$), we get $n_{\text{BG}}-\text{rank}\rho=0$ linear (`type-A') and $\text{rank}\rho/2=1$ quadratic (`type-B') Goldstone mode.

While the spin in the two valleys are in-phase and remain parallel for the Goldstone mode, they are out-of-phase in case of the second mode $p=-$; its gap, given by $g\mu_B |\vec{B}| + 2S|J_H|$, therefore remains finite when $|\vec{B}|=0$, see Fig.~\ref{fig:SI_theory}(a). This implies [cf.~Fig.~\ref{fig:SI_theory}(b)] that the $|\vec{B}|$ dependence of all microwave resonance frequencies will be linear and, for those associated with single magnon processes, of slope $g\mu_B/h$. The intercept will be zero for the lowest resonance frequency---the conventional ferromagnetic resonance mode, $f^+_{0}=E_{\vec{k}=0,p=+}/h=g\mu_B|\vec{B}|/h$---and positive for all others, including the geometric resonances $f^p_{j}=E_{\vec{k}_{j},p}/h$, where $\vec{k}_{j}$ are the discrete momenta determined by the geometry of the system. This behavior is not consistent with experiment where the lowest and dominant resonance frequency has a negative intercept.

Let us therefore next investigate the case $J_H>0$. Choosing $\vec{B}$ along the $z$-axis for concreteness, $\vec{B} = |\vec{B}|\vec{e}_z$, the classical energy (\ref{ClassicalEnergy}) is minimized by
\begin{equation}
    \vec{m}_\tau = \sin\theta_B \vec{e}_z + \tau \cos\theta_B \vec{e}_x, \quad \theta_B = \begin{cases} \arcsin{\frac{g\mu_B|\vec{B}|}{2SJ_H}}, \quad &g\mu_B|\vec{B}| \leq 2S J_H, \\
    \pi/2, \quad &g\mu_B|\vec{B}| > 2S J_H.\end{cases}
\end{equation}
Physically, this simply means that the at $\vec{B}=0$ anti-parallel spins are canted (by $\theta_B$) when we turn on $\vec{B}\neq 0$, with their ferromagnetic (anti-ferromagnetic) component along (perpendicular) to $\vec{B}$. When $g\mu_B|\vec{B}| \geq 2S J_H$, the anti-ferromagnetic component vanishes and the spins are fully aligned with the magnetic field.

\begin{figure*}
\includegraphics[width=\linewidth]{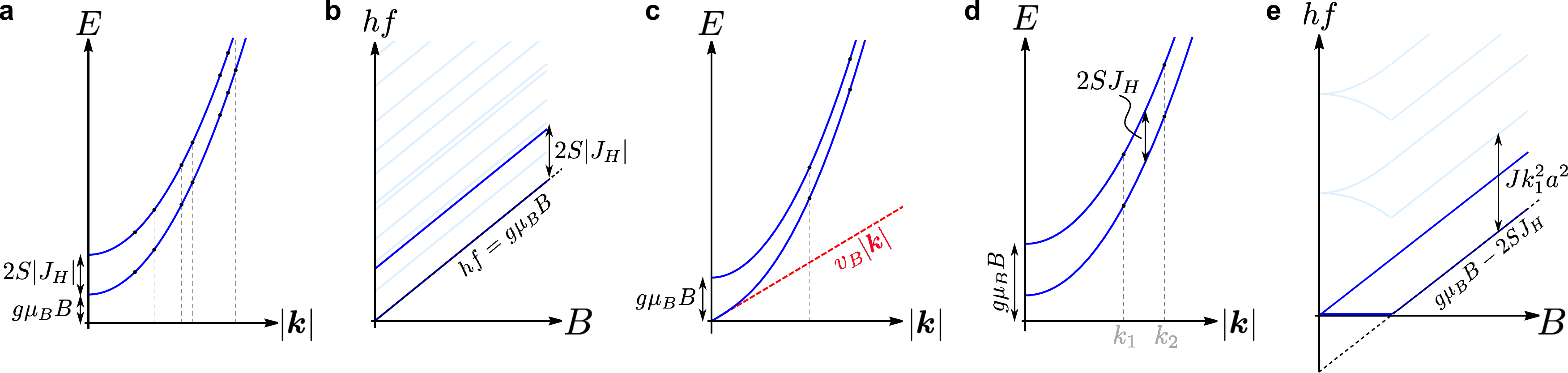}
\caption{\label{fig:SI_theory} \textbf{Magnon spectra and resonance frequencies.} Magnon spectrum (a) and resultant single-magnon resonance frequencies (b) for ferromagnetic Hund's coupling, $J_H<0$. When $J_H>0$, the magnon spectra are qualitatively different for (c) $g\mu_B |\vec{B}|<2SJ_H$ and (d) $g\mu_B |\vec{B}|>2SJ_H$, where the magnetic phases are canted anti-ferromagnets and field-aligned states, respectively. This leads to a more complex evolution of the resonance frequencies with magnetic field, as shown in (e).} 
\end{figure*}

We set $\vec{e}_{1,\tau} =-\cos\theta_B \vec{e}_z + \tau \sin \theta_B \vec{e}_x$ and $\vec{e}_{2,\tau}=\tau \vec{e}_y$, which yields $t=-S J_H \sin^2\theta_B$ and $\Delta = SJ_H \cos^2\theta_B$. The resultant form of $\hat{H}_{H}$ can be diagonalized by introducing new bosons $\hat{\gamma}'_{\vec{k}}$ related to $b_{\vec{k},\tau}$ by a Bogoliubov transformation yielding
\begin{subequations}\begin{align}
    \hat{H}_{H} &= \text{const.} + \sum_{\vec{k},p=\pm} E'_{\vec{k},p} \hat{\gamma}'^\dagger_{\vec{k},p}\hat{\gamma}'^\pdagger_{\vec{k},p} + \mathcal{O}(1/S^0),  \\
    E'_{\vec{k},p} &= \sqrt{(\xi_{\vec{k}} + S J_H \cos 2\theta_B + g\mu_B |\vec{B}|\sin \theta_B + p\, S J_H \sin^2 \theta_B)^2 - \left(S J_H \cos^2\theta_B\right)^2}. \label{DiagonalizedModel}
\end{align}\end{subequations}
To understand the magnon spectra, let us first focus on the canted antiferromagnetic regime, $0< g\mu_B|\vec{B}|<2S J_H$, where $2S J_H \sin\theta_B = g\mu_B |\vec{B}|$ and we get
\begin{subequations}\begin{align}
    E'_{\vec{k},-} &= \sqrt{\xi_{\vec{k}}(\xi_{\vec{k}}+ 2S J_H \cos^2\theta_B)} = v_B |\vec{k}| + \mathcal{O}(|\vec{k}|^3), \quad v_B = \sqrt{2S J_H Ja^2} |\cos \theta_B|, \label{ExpressionForVelocity} \\
    E'_{\vec{k},+} &= \sqrt{\xi_{\vec{k}}(\xi_{\vec{k}}+2SJ_H(\cos 2\theta_B+ 3\sin^2\theta_B)) + (2 S J_H \sin\theta_B)^2} = \Delta_B  + \mathcal{O}(|\vec{k}|^2), \quad  \Delta_B = g\mu_B |\vec{B}|.
\end{align}\label{DispersionRegime1}\end{subequations}
As also illustrated in Fig.~\ref{fig:SI_theory}(c), we find a gapped mode ($p=+$) with gap $\Delta_B$ given by the Zeeman energy and a gapless mode ($p=-$) that has a linear dispersion. The latter is the single type-A \cite{CountingGoldstones} Goldstone mode of the problem (note the ground state breaks the single continuous symmetry, given by spin rotations along the magnetic field axis, such that $n_{\text{BG}}=1$ and $\text{rank}\rho =0$, leading to $n_{\text{BG}}-\text{rank}\rho=1$ type-A and $\text{rank}\rho/2=0$ type-B Goldstone modes). The vanishing of $\Delta_B$ for $\vec{B}=0$ is also expected since the ground state breaks two out of the three generators $\hat{\Lambda}_i$, however, as opposed to the ferromagnet we have $\rho=0$, leading to two (zero) type-A (type-B) Goldstone modes.

For $g\mu_B|\vec{B}|>2S J_H$, the spins in the two valleys are fully aligned such that $n_{\text{BG}}=\text{rank}\rho=0$ and no Goldstone modes are expected. This is indeed reproduced from \equref{DiagonalizedModel} which can be written in this limit as
\begin{equation}
    E'_{\vec{k},p} = \xi_{\vec{k}} + g\mu_B |\vec{B}| + (p-1) S J_H.  \label{DispersionRegime2}
\end{equation}
We see from this expression and in Fig.~\ref{fig:SI_theory}(d) that both modes have a finite gap, given by $\Delta_p = g\mu_B |\vec{B}| + (p-1) S J_H$.

Taken together, \equsref{DispersionRegime1}{DispersionRegime2} show that the $|\vec{B}|$ dependence of the lowest resonance frequency $f^-_{0}=E_{\vec{k}=0,p=-}/h$ is given by a straight line with slope $g\mu_B/h$ and negative intercept $-2SJ_H/h$ for $g\mu_B|\vec{B}|>2S J_H$ (and zero below that field strength) while $f^+_{0}=E_{\vec{k}=0,p=+}/h=g\mu_B |\vec{B}|$ has vanishing intercept, see dark blue lines in Fig.~\ref{fig:SI_theory}(e). These two features are in excellent agreement with experiment (resonance mode $r_1$ and $r_2$) and we can use the measured intercept (from mode $r_1$) to extract the value of $J_H$. Beyond the existence of a set of collective bulk modes with the expected magnetic field dependence, it is an important question to understand the precise mechanism by which these modes are excited by microwaves  in the setup and how this translates to a transport response. We leave a detailed theoretical and systematic experimental study of this to future work, but note that, potentially, the reduced symmetries at samples edges and due to the proximate WSe$_2$ layer might play an important role.

To also capture the higher resonance modes at least qualitatively, we make a continuum/low-momentum approximation, $\xi_{\vec{k}} \sim J \vec{k}^2 a^2$, take the system geometry to be rectangular (of size $L_x \times L_y$), and quantize the momenta as $\vec{k}=2\pi (n/L_x,m/L_y)^T$, $n,m\in\mathbb{Z}$. The first few resonance modes are indicated in light blue in the schematics in Fig.~\ref{fig:SI_theory}(e). In the main text, we use this description to fit the third resonance mode (denoted by $r_3$), which allows us to extract $J$ and, together with $J_H$, also the velocity $v_B$ in \equref{ExpressionForVelocity}; without further fitting parameters, this model also captures the resonance modes $r_3$, $r_4$, and, as already mentioned, $r_2$.

\subsection{Other Dirac revivals}

Finally, we come back to the remaining orders in \equref{CandidateOrders} and provide arguments why they are less natural candidates than the first two discussed at length above. We first note that Ref.~\cite{LakePairingPhen} recently argued, based on a detailed analysis of the experimental literature on twisted bilayer and trilayer graphene, that only U(1)$_v$-symmetry-preserving, spin-polarized states with zero momentum are consistent with experiment, leaving only the first two options in \equref{CandidateOrders}. Moreover, also our microwave data is less naturally consistent with the remaining options. First, the last two states in \equref{CandidateOrders} break U(1)$_v$ and will hence exhibit a Goldstone mode that is not gapped out by the magnetic field. However, all resonance frequencies we observe increase linearly with slope $g\mu_B/h$. Second, also $\mu_{x,y}\vec{s}$ and its Hund's partner $\tau_z\mu_{x,y}\vec{s}$ will exhibit a gapless Goldstone mode in the presence of a magnetic field. To see this, let us write the coupling to the low-energy flat-band fermions (with operators $f_{\vec{q}}$) as
\begin{equation}
    \Delta H = \sum_{\vec{q}}\sum_{j=x,y}\sum_{k=x,y,z}\sum_{v=\pm} \phi^v_{j,k} f^\dagger_{\vec{q}} \mu_j s_k P_v f^\pdagger_{\vec{q}} = \frac{1}{2}\sum_{\vec{q}}\sum_{v=\pm}  f^\dagger_{\vec{q}} (\mu_x+i\mu_y) \vec{\varphi}_v \cdot\vec{s} P_v f^\pdagger_{\vec{q}} + \text{H.c.},
\end{equation}
where $P_\pm=(1\pm\tau_z)/2$ and the sum over $\vec{q}$ is restricted to the vicinity of the Dirac cones; we can see that the underlying order is described either by two real $2\times 3$-matrix-valued order parameters $\phi^v_{j,k}$---one for each valley---or by two complex three-component vector order parameters $\vec{\varphi}_v$ [$(\vec{\varphi}_v)_k= \phi^v_{1,k} - i\, \phi^v_{2,k}$]. 
Using the latter, spin-rotation acts like the corresponding SO(3) rotation of the vector $\vec{\varphi}_\pm$, an elementary moir\'e translation corresponds to $\vec{\varphi}_\pm \rightarrow \omega \vec{\varphi}_\pm$, $\omega = e^{\frac{2\pi i}{3}}$, time-reversal acts as $\vec{\varphi}_\pm\rightarrow -\vec{\varphi}_\mp$, and two-fold rotation $\vec{\varphi}_\pm\rightarrow \vec{\varphi}^*_\mp$. The leading-order coupling of the Zeeman field to the order parameter consistent with these symmetries reads as $\Delta \mathcal{F} = i g \sum_{v=\pm} v \vec{\varphi}_v^* \times \vec{\varphi}^{\phantom{*}}_v \cdot \vec{B}$. In a finite magnetic field, say $\vec{B} = |\vec{B}| \hat{\vec{e}}_z$, we thus expect $\vec{\varphi}_\pm \propto (1,\pm i\, \text{sgn}(g),0)^T$ corresponding to
\begin{eqnarray}
    \Delta H \propto \sum_{\vec{q}}\sum_{v=\pm} f^\dagger_{\vec{g}} P_v \left( \mu_x s_x - v \,\text{sgn}(g) \mu_y s_y\right) f^\pdagger_{\vec{q}}.
\end{eqnarray}
This state breaks one of the two generators of the continuous symmetry group U(1)$_v\times$SO(2)$_{z}$ in the presence of a magnetic field ($n_{\text{BG}}=1$ and $\text{rank}\rho =0$), leading to a single (type-A) Goldstone mode \cite{CountingGoldstones}. As already mentioned above, we find no indications of such a mode and its associated geometric resonances. Taken together, our findings are most naturally consistent with $\tau_z\vec{s}$ in \equref{CandidateOrders}.

\subsection{Discussion of other possible origins}
For completeness, we will here discuss a few other possible microscopic origins of the observed microwave resonances. However, as we will see, none of them can explain the experimental phenomenology as well as the antiferromagnetic spin polarization studied in detail above.

First, one might wonder whether ferromagnetic spin polarization across the two valleys, i.e., the first candidate order in \equref{CandidateOrders}, supplemented with spin-orbit coupling can give rise to a ferromagnetic resonance mode with negative intercept, as observed in experiment (mode $r_1$). To study the associated possible ferromagnetic resonances, we have applied the formalism of Refs.~\cite{ResFreq1,ResFreq2} using the free-energy expression $\mathcal{F} \sim -\vec{S}\cdot \vec{H} + \alpha \, S_z^2 + \beta_1 (S_x^2+S_y^2)S_z^2 + \beta_2 \, S_z^4 + \beta_3 \, S_x(S_x^2-3S_y^2)S_z$, where $\vec{S}=(S_x,S_y,S_z)^T$ is a three-component unit vector pointing along the direction of the system's magnetization. It is clear by symmetry that $\alpha$, $\beta_{1,2,3}$ can only be non-zero if spin-orbit coupling is finite. We studied the resulting behavior of the ferromagnetic resonance frequency as a function of magnetic field for different $\alpha$, $\beta_{1,2,3}$. We found that, while it is possible to find parameters to get a large-field regime where the resonance frequency grows approximately linearly in field with negative offset for an in-plane (out-of-plane) magnetic field, rotating the field out-of-plane (in-plane) for the same parameters changes the resonance frequencies significantly. This is not consistent with experiment. 

Another possibility is that the resonances are associated with particle-hole excitations across a gap that grows linearly and with slope $g\mu_B$ with magnetic field; as such, it must be a gap that separates states with spin polarizations parallel and anti-parallel to the external magnetic field.
Given the onset of resonances at $|\nu|=2$, it might be natural to conclude that this is the gap of the Dirac revival itself, i.e., between those bands pushed below and those at the Fermi level. However, we expect that this gap is of the order of the bandwidth of the flat bands and, hence, significantly larger than the energies where we observe resonances (of order of $10\,\textrm{GHz}$ which is about $0.04\,\textrm{meV}$). 
Another related possibility is that the resonances are across a gap induced by spin-orbit coupling---either intrinsic to graphene or due to the nearby WSe$_2$ layer---once the interaction-induced Dirac-revival has taken place. While it might be possible that the band reconstruction due to the Dirac revival enhances the visibility of microwave resonances in transport, potentially providing a route to understand the onset at $|\nu|=2$, we obtain the same problem as for ferromagnetic resonances and spin-orbit coupling: the resonance frequency behavior with magnetic field would generically differ for in- and out-of-plane magnetic fields, inconsistent with our measurements. Finally, note that the noninteracting picture based on spin-orbit coupling put forward by previous works on microwave resonance in MLG~\cite{Sichau2019ESR,Blick2020} does not provide a natural interpretation of our findings either: not only does the same reason (spin-orbit coupling generically leading to different resonances for in- and out-of-plane fields) apply here too, but also a noninteracting description is very unlikely able to capture the absence of microwave resonance at the CNP and its onset right at $|\nu|=2$.

\section{Materials and Methods}

\subsection{Device Fabrication}

{\bf{Device fabrication and measurement}} The vdW assembly procedures for creating the tBLG/\WSe\ heterostructure is detailed as the following: Each layer of the two-dimensional material is exfoliated onto a silicon chip, which is then picked up sequentially with a PC/PDMS stamp. The monolayer graphene is cut in two halves using an AFM tip. The two pieces are picked up with an intended rotational misalignment of $1.2^\circ$, slightly larger than the final twist angle of the device. 
The hBN substrate and tBLG are misaligned with an angle of $15^\circ$, whereas tBLG and \WSe\ are rotationally misaligned by $44^\circ$, which is equivalent to $16^\circ$. We note that a twist angle of $16^\circ$ between tBLG and \WSe\ is expected to give rise to maximum SOC strength.
Independent controls on charge carrier density $n_{tBLG}$ and displacement field $D$ are achieved by applying D.C. voltage to the top and bottom graphite gate electrodes, $V_{top}$ and $V_{bot}$, respectively.

All layers of two-dimensional materials used in the device are produced using the mechanical exfoliation method, which are subsequently stacked together by a poly(bisphenol A carbonate) (PC)/polydimethylsiloxane (PDMS) stamp. The tBLG is assembled using the ``cut-and-stack" technique~\cite{Saito2019decoupling}, in which a monolayer graphene is cut in half using an atomic force microscope (AFM) before being picked up to improve the twist angle accuracy and homogeneity. The layers of the device are composed of (from top to bottom): BN (36 nm), graphite (7 nm), BN (61 nm), \wse (2 nm), tBLG, BN (37 nm), graphite (5 nm). The thickness of the layers are measured with an AFM. 

The fabrication of the device follows the standard electron-beam lithography, reactive-ion etching (RIE) and electron-beam evaporation procedures. First the top graphite is removed from the contact region using CHF$_3$/O$_2$ plasma in the RIE, then the same recipe is used to define the Hall-bar shape and expose the graphene edge for the contact, finally Cr/Au (2/100nm) is deposited to form the electrodes for the tBLG and both graphite gates.

An important feature of the heterostructure used here is the atomic interface between \WSe\ and MATBG. This interface is shown to induce strong coupling between the spin and orbital degrees of freedom in MATBG ~\cite{Lin2021SOC,Island2019spin,Koshino2019SOC}. The idea is that the presence of SOC will convert a resonance response in the spin channel into changes in the sample resistivity, which can be detected through resistance measurement.

\subsection{Transport Measurement}

The dual-gated structure allows independent control of carrier density in the tBLG, $n_{tBLG}$, as well as displacement field $D$ in the out-of-plane direction. Such control is achieved by applying a DC gate voltage to top and bottom graphite electrodes, $V_{top}$ and $V_{bot}$, respectively. $n_{tBLG}$ and $D$ can be obtained using the following equations: 
\begin{eqnarray}
n_{tBLG} &=& (C_{top}V_{top}+C_{bot}V_{bot})/e+n^0_{tBLG}, \label{EqM3}\\
D&=& (C_{top}V_{top}-C_{bot}V_{bot})/2\epsilon_0, \label{EqM4}
\end{eqnarray} 
\noindent
where $C_{top}$ ($C_{bot}$) is the geometric capacitance between top (bottom) graphite and tBLG, and is determined from the conventional Hall resistance. $n^0_{tBLG}$ is the intrinsic doping in tBLG. 
In all of the measurements detailed above, $D$ is constant at $200$ mV/nm. 

Standard low frequency lock-in techniques with Stanford Research SR830 amplifier are used to measure resistance $R_{xx}$ and $R_{xy}$, with an excitation current of $10$ nA at a frequency of $17.77$ Hz. 
The parallel field measurement is performed by mounting the device on a homemade adapter which fixes the device at a desired angle relative to the field orientation. 
The atomically-thin \wse\ has a large band gap and thus becomes insulating at cryogenic temperature. Therefore, it can be viewed as part of the dielectric layer, and does not contribute to the electrical transport signal directly \cite{Island2019spin}.

For the microwave measurements shown in the main text, longitudinal resistance \Rxx\ is measured when sweeping the external magnetic field, either in the in-plane or out-of-plane orientation.  In the case of the $f-B$ data, $B$-field sweep is performed with fixed MW frequency, \emph{i.e.} $B$ is the fast axis, and MW frequency is the slow axis. 
Along the same vein, $B$ is again the fast axis and moir\'e filling $\nu$ the slow axis for the $\nu-B$ maps. In this case, MW power and frequency are fixed at constant values for the entire measurement. To eliminate the slow varying $B$-dependent background in \Rxx, a high-pass filter is added in
post-measurement analysis for each $B$-sweep line. The cutoff frequency of the high-pass filter, which is in arbitrary units that only depends on the density of $B$ points, is chosen in a way that it does not interfere with microwave-induced resonance features. 

\begin{figure*}
\includegraphics[width=0.65 \linewidth]{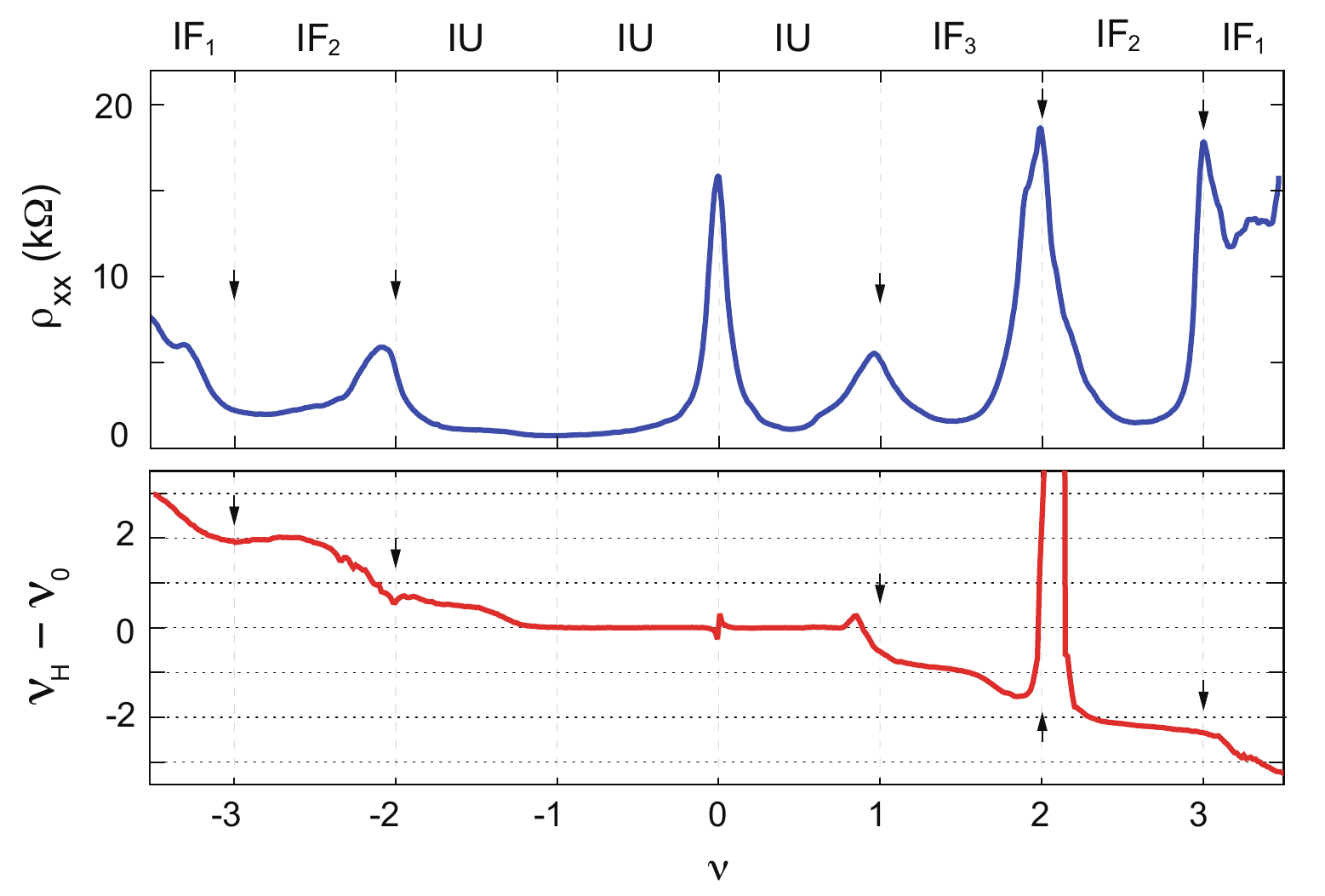}
\caption{\label{fig:Hall} {\bf{Identification of density regimes with different isospin order.}} Longitudinal resistance \Rxx\ (top panel) and renormalized Hall density (bottom panel) as a function of moir\'e filling across the flatband. Vertical black arrows mark isospin polarization transitions, defined by the peak in \Rxx\ and a step in Hall density. The identification of different isospin order is based on the main sequence of quantum oscillation as shown in Fig.~\ref{fig:LLhighB}, which follows the same criterion as in Ref.~\cite{Saito2021pomeranchuk,Liu2022DtTLG,Lin2021SOC}.
}
\end{figure*}

\begin{figure*}
\includegraphics[width=0.7 \linewidth]{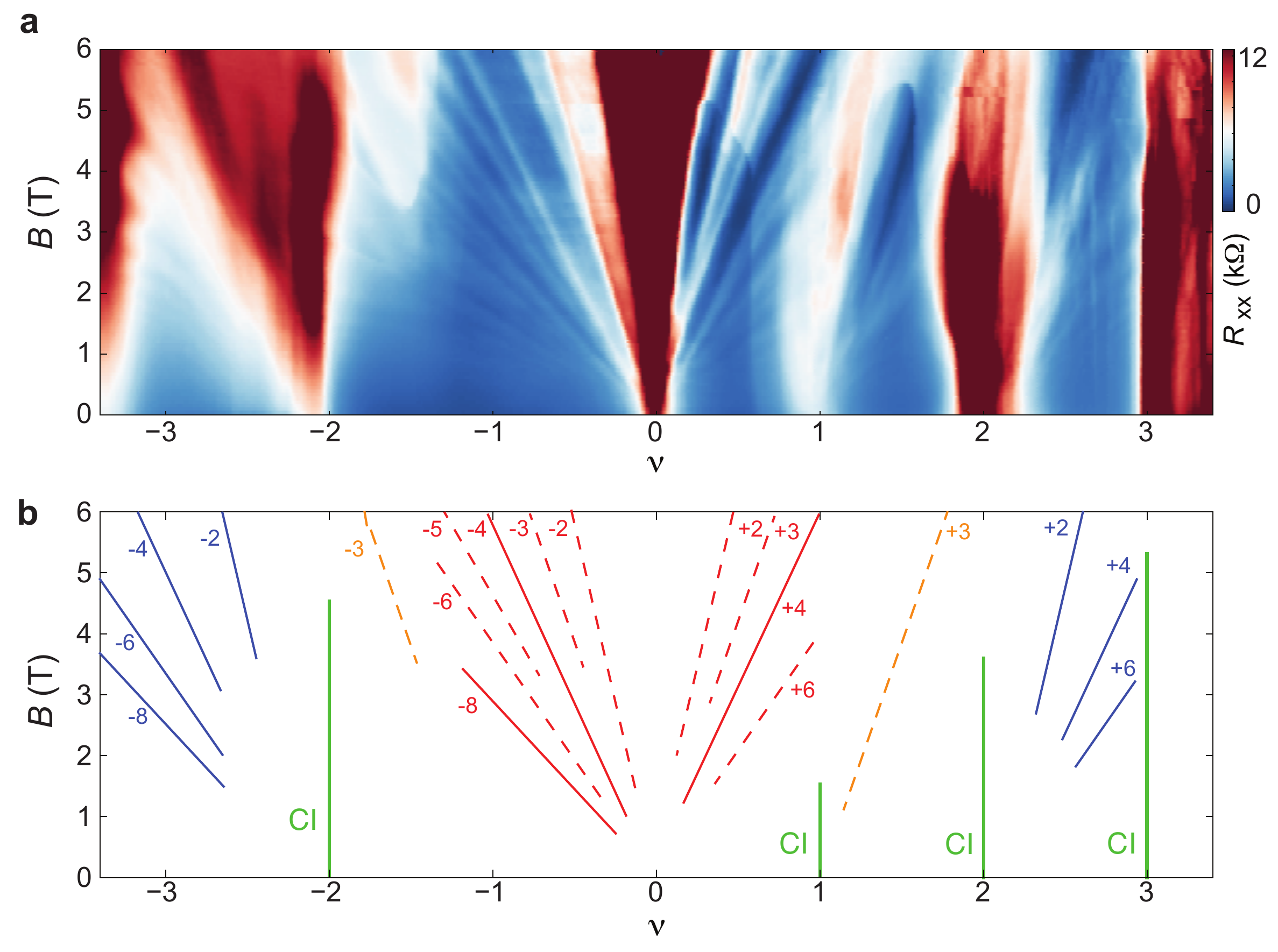}
\caption{\label{fig:LLhighB} {\bf{Magneto-transport measurement across the moir\'e flat band.}} (a) Longitudinal resistance \Rxx\ as a function of $B$ and moir\'e filling $\nu$. Quantum oscillations are observed to emanate from integer moir\'e fillings where the Fermi surface is reconstructed by the spontaneous isospin polarization.  (b) Schematic labelling the most prominent states in panel (a). Vertical green lines denote correlation-driven insulators (CIs), which correspond to the resistance peaks in Fig.~\ref{figN}b and Fig.~\ref{fig:Hall}. The isospin degeneracy of the underlying Fermi surface can be identified based on the main sequence of quantum oscillations. For instance, IF$_2$ on both electron and hole side exhibits a main sequence of $\nu_{LL}=2$, $4$, $6$..., which is in excellent agreement with a 2-fold degeneracy. Similarly, IF$_3$ corresponds to a 3-fold degeneracy. 
}
\end{figure*}

\begin{figure*}
\includegraphics[width=0.45 \linewidth]{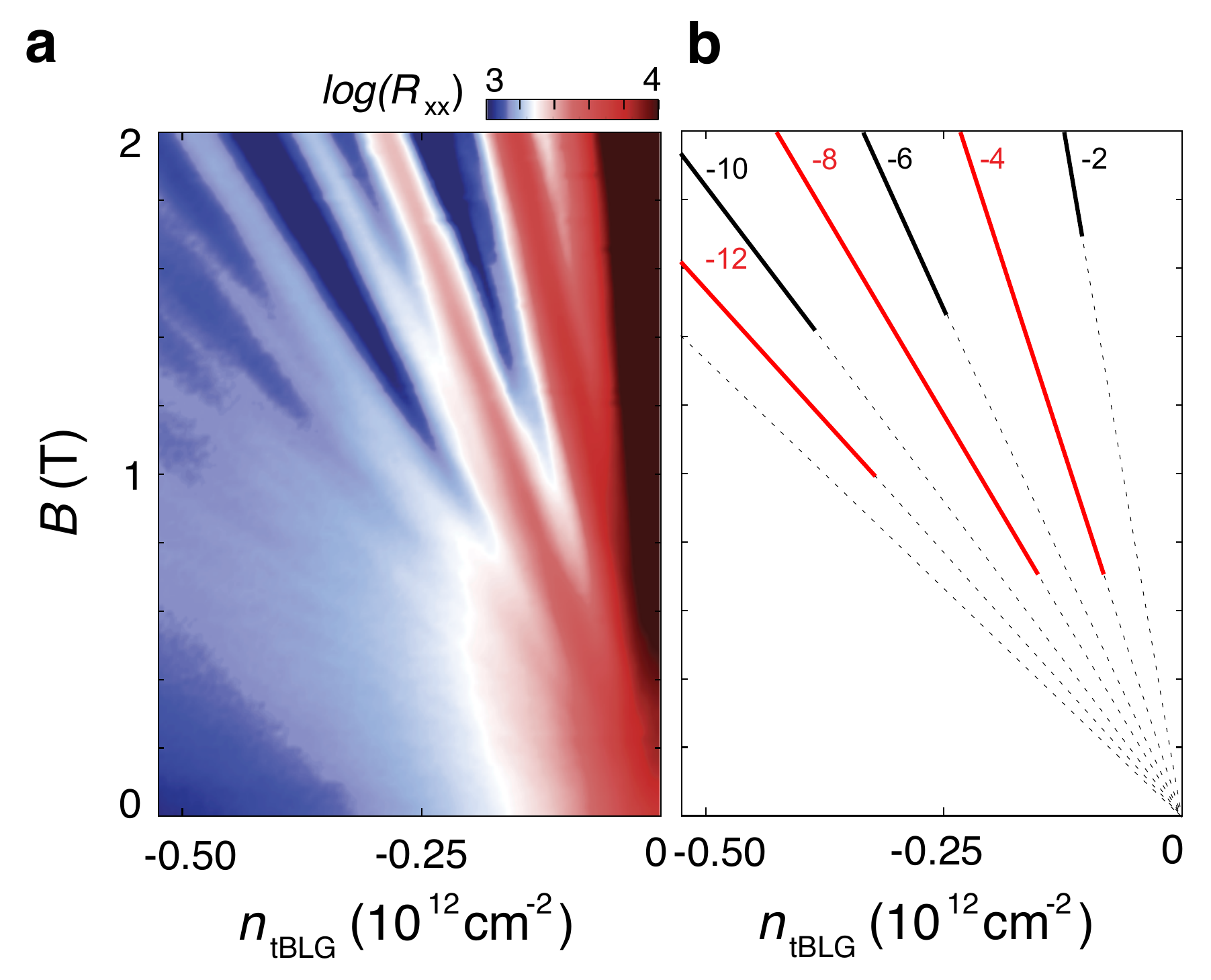}
\caption{\label{fig:LLlowB} {\bf{Quantum oscillation near the CNP at low magnetic field.}} (a) Longitudinal resistance \Rxx\ as a function of $B$ and carrier density $n$. (b) Schematic labelling the most prominent quantum Hall effect (QHE) states emanating from the CNP. The most robust QHE features that develops below $B < 1$ are marked with red solid lines, which correspond to QHE states that preserve the degeneracy of the underlying Fermi surface. The sequence of $\nu_{LL}=4$, $8$, $12$ signify a 4-fold degenerate Fermi surface near the CNP, which is consistent with the indentification of the IU regime in Fig.~\ref{figN}.
}
\end{figure*}

\begin{figure*}
\includegraphics[width=0.7\linewidth]{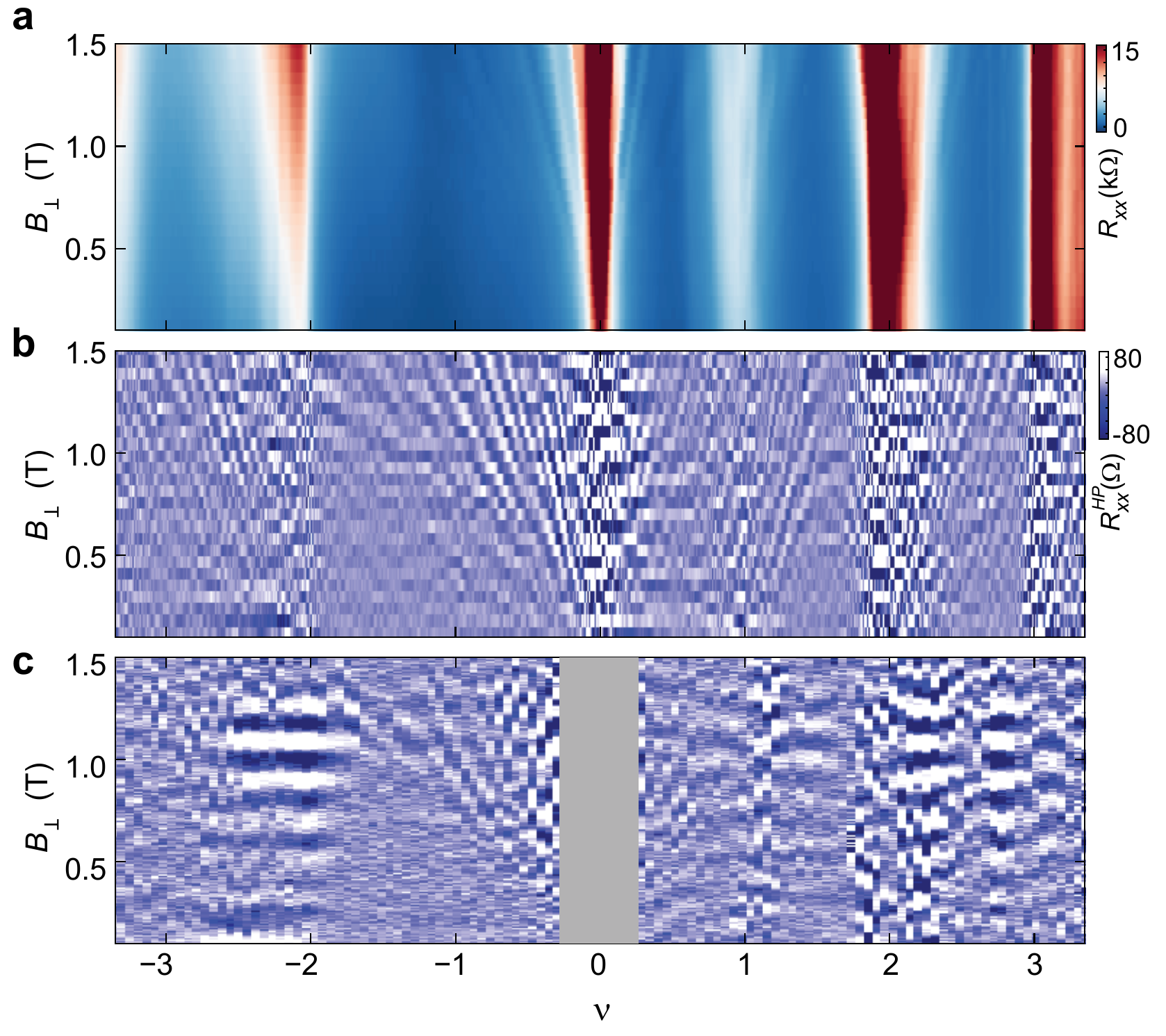}
\caption{\label{fig:LL} {\bf{Landau fan with and without microwave radiation}} (a) Longitudinal resistance \Rxx\ as a function of out-of-plane magnetic field \Bperp\ and moir\'e filling $\nu$ with no microwave signal. (b) High-pass filter of the no MW Landau fan shown in (a) where the fan features are clear across the moir\'e range and there are no signatures of MW resonance as compared to Fig. \ref{figN}c. (c) \RHP\ as a function of moir\'e filling and \Bperp\ measured with microwave radiation. This is the same map as shown in Fig.~\ref{figN}c, but with an extremely saturated color scale. This reveals the weaker resonance response in the density regime of IF$_3$, as well as IF$_1$ on the electron doping side of the moir\'e band. Even with the saturated color scale, no resonance mode is observed in the IU regime, confirming the observation in Fig.~\ref{figN}. Similarly, no resonance mode is observed in the IF$_1$ regime on the hole doping side. Dirac revival at $\nu=-3$ is not fully developed, evidenced by the absence of a resistance peak. As such, the absence of a resonance response associated with IF$_1$ is consistent with the proposed connection between resonance response and Dirac revival. 
We note that there exists an alternative possibility that $r_2$ through $r_4$ all result from the geometric resonance of the magnon mode. This scenario will introduce slight modification to the estimated value of the spin stiffness $J$. Most importantly, the value of $J_H$ will remain the same, since the intercept of the $r_1$ mode is unchanged.  \ Our findings suggest that a more accurate determination of $J$ and $v_B$ of the moir\'e flatband can be achieved by electron spin resonance measurement in a sample with well-defined cavity, such as the one used to study quantum Hall magnons ~\cite{Fu2021magnon}. 
It is worth pointing out that  quantum Hall effect states around the CNP remain robust in the presence of microwave radiation (Fig.~\ref{figN}c-d and panel (c) here). 
The energy gap associated with Landau levels, which is proportional to \vecB, diminishes with decreasing \vecB. As a result, quantum Hall effect states are susceptible to small changes in the temperature in the low-\vecB\ regime. The fact that microwave radiation does not affect quantum Hall effect states at low-$\vec{B}$ offers strong evidence that the impact on electron temperature from microwave radiation  is minimal and inconsequential. This observation further validates that the resonance behavior originates from collective excitations of the moir\'e flatband. 
The CNP of MATBG becomes highly resistive in the presence of an out-of-plane magnetic field. This gives rise to an abundance of spurious signals in longitudinal resistance even in the absence of microwave radiation (panel b). As such, we mask the CNP of the $\nu-B$ map (panel c and Fig.~\ref{figN}c).
}
\end{figure*}

\begin{figure*}
\includegraphics[width=0.6\linewidth]{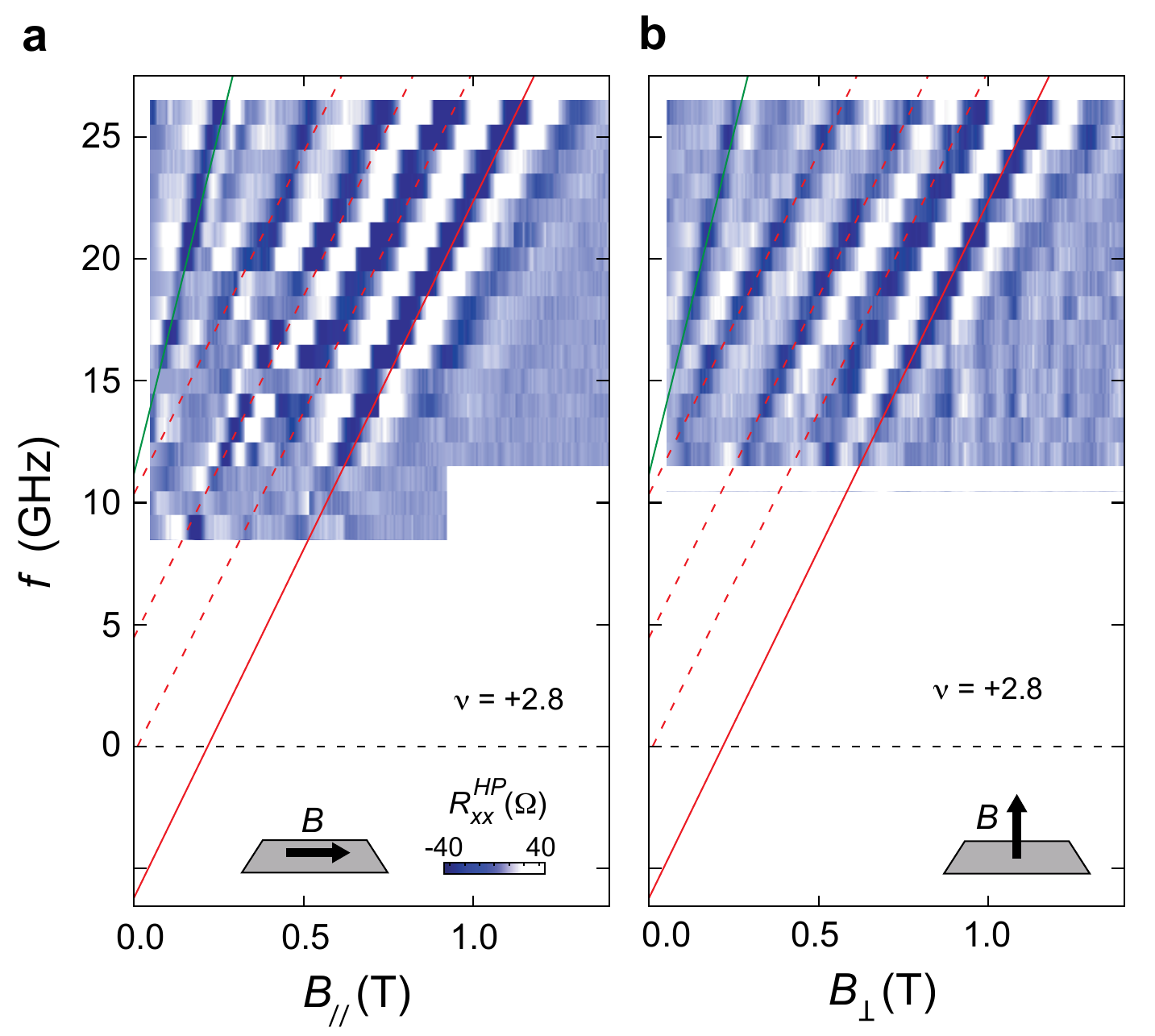}
\caption{\label{fig:Bdirection} {\bf{Dependence on the orientation of $B$ at $\boldsymbol{\nu = +2.8}$}} (a) \RHP\ as a function of $\nu$ and (a) \Bpara, (b) \Bperp\ measured at $\nu=+2.8$ and $T = 50$ mK. All five resonance modes remain mostly at the same location with different $B$-field orientation.
}
\end{figure*}

\begin{figure*}
\includegraphics[width=0.7\linewidth]{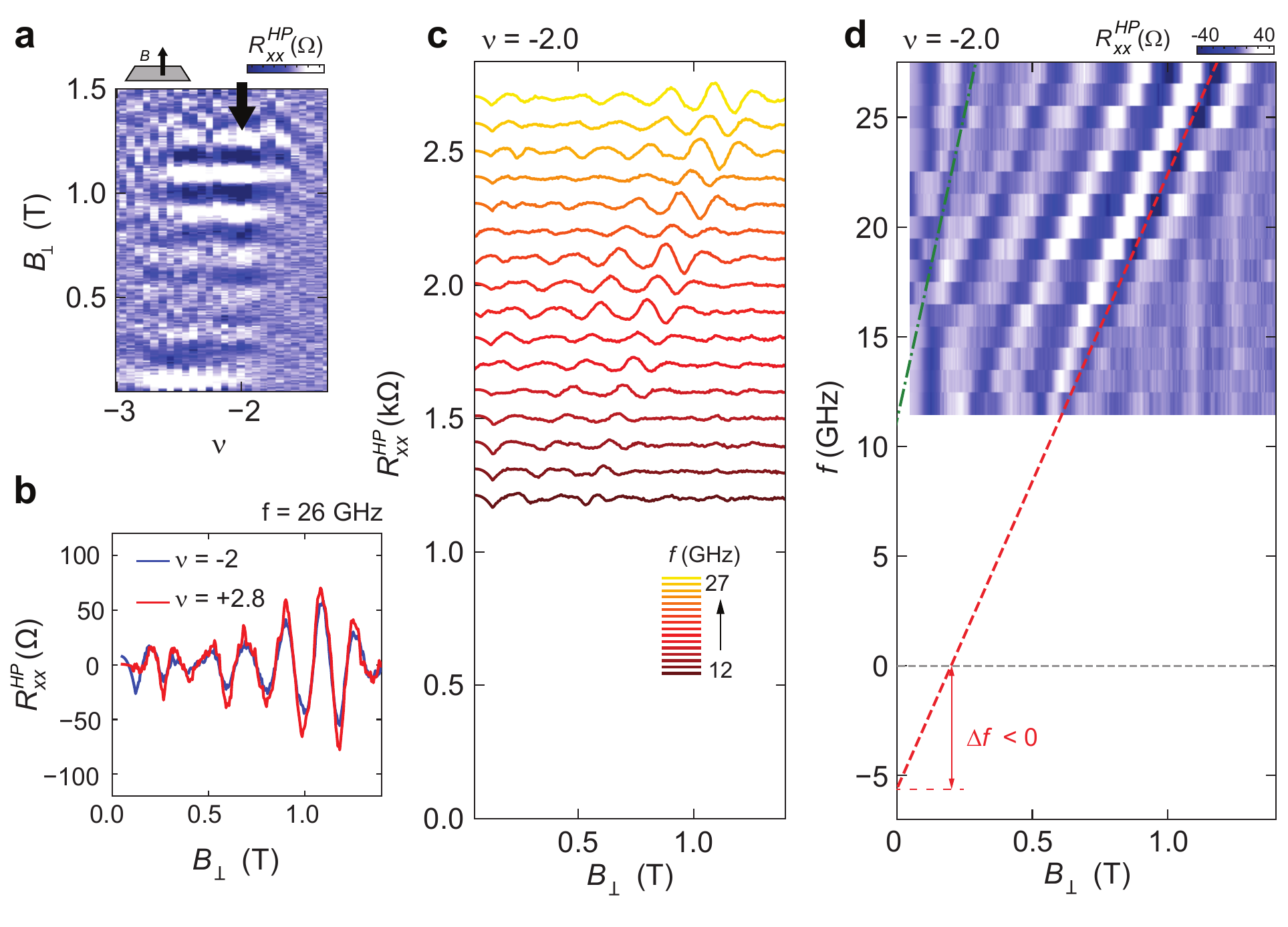}
\caption{\label{fig:SI_perp_density} {\bf{Frequency dependence of $\nu = -2$ with out-of-plane field}} (a) \RHP\ as a function of $\nu$ and \Bperp\ as shown in Fig. \ref{figN}c, focusing on the moir\'e filling range near $\nu=-2$. Microwave frequency dependence near $\nu$ = -2 is shown in (c) a the waterfall plot of \RHP, and (d) a \RHP\ map with the same $g=2$ and $g=4$ resonance modes as described in Fig. \ref{fig2}b for $\nu = +3$. The negative intercept at $B = 0$ in (d) also indicates antiferromagnetic coupling for $\nu = -2$, in addition to $+2.8$. 
}
\end{figure*}

\end{widetext}

\end{document}